\documentclass[12pt]{article}

\usepackage[dvipsnames]{xcolor}
\usepackage{amsmath}
\usepackage{amssymb}
\usepackage{dsfont}
\usepackage[english]{babel}
\usepackage[utf8]{inputenc}
\usepackage{times}
\usepackage{graphics}
\usepackage{graphicx}
\usepackage[T1]{fontenc}
\usepackage[official,right]{eurosym}
\usepackage{url}
\usepackage{eso-pic}

\title{Fairness and Explainability in Automatic Decision-Making Systems. \\ A challenge for computer science and law}
\author{Thierry Kirat$^\dag$, Olivia Tambou$^*$, \\ Virginie Do$^\ddag$, Alexis Tsoukiàs$^\ddag$, \\ $^\ddag$CNRS-IRISSO, PSL, Université Paris Dauphine \\ $^*$CR2D, PSL, Université Paris Dauphine \\ $^\ddag$CNRS-LAMSADE, PSL, Université Paris Dauphine}
\date{}

\begin{document}

\thispagestyle{empty}

\enlargethispage*{8cm}
 \vspace*{-38mm}

\AddToShipoutPictureBG*{\includegraphics[width=\paperwidth,height=\paperheight]{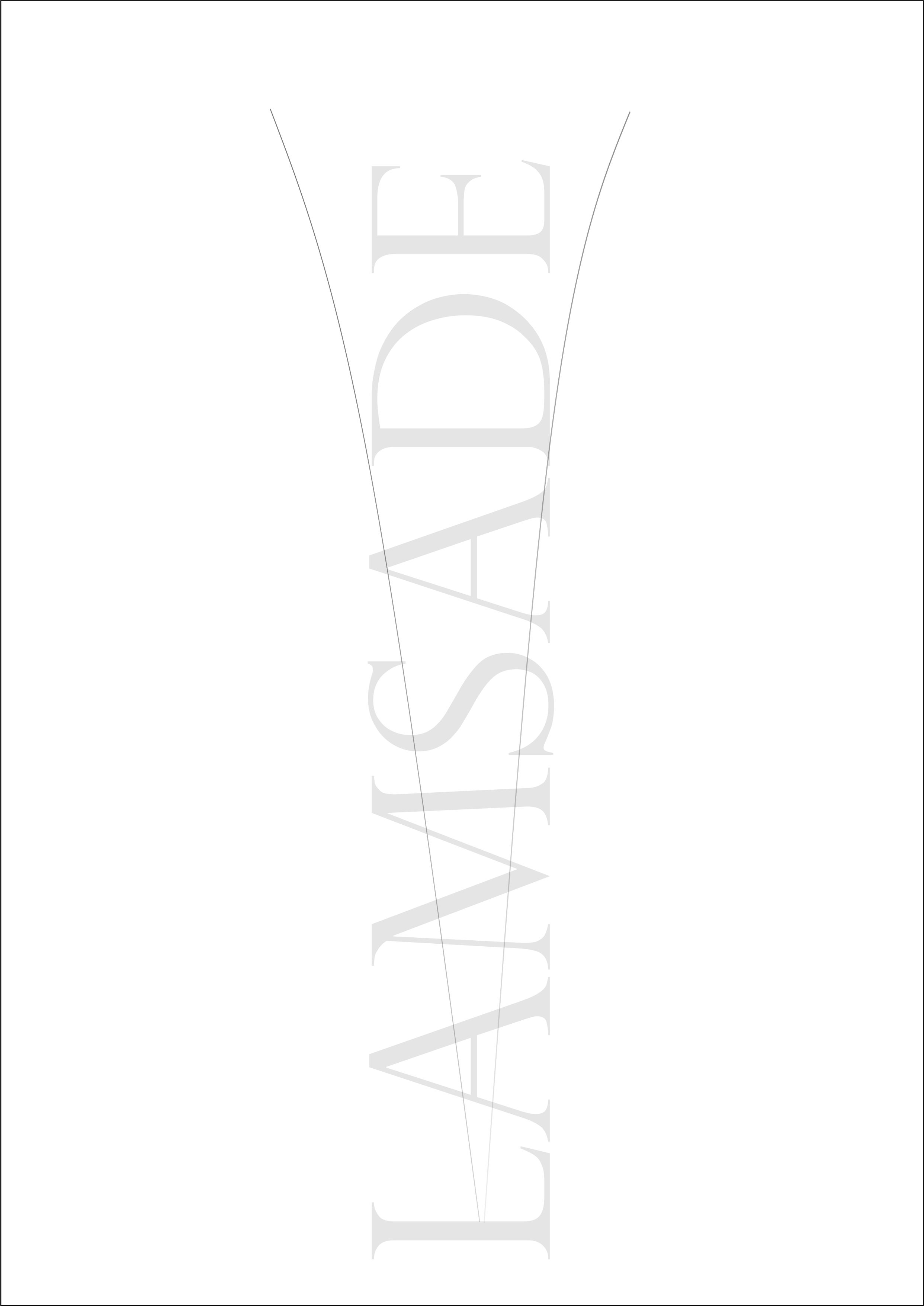}}

\begin{minipage}{24cm}
 \hspace*{-28mm}
\begin{picture}(500,700)\thicklines
 \put(60,670){\makebox(0,0){\scalebox{0.7}{\includegraphics{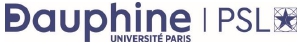}}}}
 \put(60,70){\makebox(0,0){\scalebox{0.3}{\includegraphics{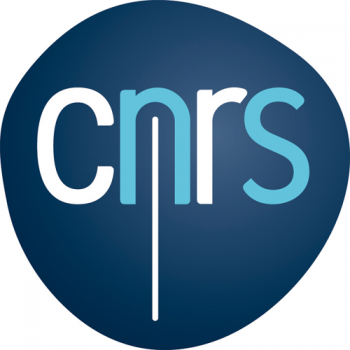}}}}
 \put(320,350){\makebox(0,0){\Huge{CAHIER DU \textcolor{BurntOrange}{LAMSADE}}}}
 \put(140,10){\textcolor{BurntOrange}{\line(0,1){680}}}
 \put(190,330){\line(1,0){263}}
 \put(320,310){\makebox(0,0){\Huge{\emph{402}}}}
 \put(320,290){\makebox(0,0){May 2022}}
 \put(320,210){\makebox(0,0){\Large{Fairness and Explainability}}}
 \put(320,190){\makebox(0,0){\Large{in Automatic Decision-Making Systems.}}} \put(320,170){\makebox(0,0){\Large{A challenge for computer science and law}}}
 \put(320,100){\makebox(0,0){\Large{Thierry Kirat, Olivia Tambou, Virginie Do}}}
 \put(320,80){\makebox(0,0){\Large{Alexis Tsoukiàs}}}
 \put(320,670){\makebox(0,0){\Large{\emph{Laboratoire d'Analyse et Mod\'elisation}}}}
 \put(320,650){\makebox(0,0){\Large{\emph{de Syst\`emes pour l'Aide \`a la D\'ecision}}}}
 \put(320,630){\makebox(0,0){\Large{\emph{UMR 7243}}}}
\end{picture}
\end{minipage}

\newpage

\addtocounter{page}{-1}

\maketitle

\begin{abstract}

The paper offers a contribution to the interdisciplinary constructs of analyzing fairness issues in automatic algorithmic decisions. Section 1 shows that technical choices in supervised learning have social implications that need to be considered. Section 2 proposes a contextual approach to the issue of unintended group discrimination, i.e. decision rules that are facially neutral but generate disproportionate impacts across social groups (e.g., gender, race or ethnicity). The contextualization will focus on the legal systems of the United States on the one hand and Europe on the other. In particular, legislation and case law tend to promote different standards of fairness on both sides of the Atlantic. Section 3 is devoted to the explainability of algorithmic decisions; it will confront and attempt to cross-reference legal concepts (in European and French law) with technical concepts and will highlight the plurality, even polysemy, of European and French legal texts relating to the explicability of algorithmic decisions. The conclusion proposes directions for further research.

\end{abstract}

\section{Introduction}

The fairness of algorithmic decisions based on machine learning has been a fundamental issue in computer science research for more than two decades; more recently, it has been the object of increasingly multidisciplinary research, to which legal specialists and social scientists contribute.
We will develop this point in the text, but for the moment it is important to underline that what was initially a societal issue treated in a technical way, without the need for an external viewpoint, has become the object of a collaboration that can be described as fruitful, between computer scientists and researchers in law and public policy analysis.

This article focuses on the conditions for horizontal fairness among individuals affected by automated decisions based on learning algorithms. Without ignoring the technical possibilities of achieving fair machine learning algorithms (\cite{Barocasetal2019}, \cite{Chouldechova2017}, \cite{CorbettGoel2018}, \cite{Kleinbergetal2016}), we argue that the goal of fairness:

\begin{itemize}
 \item[a)] Cannot be limited to an exclusively technical issue, i.e. algorithm and model design without reference to other than technical dimensions,
 \item[b)] Supposes that social choices are made to define the desirable form of equity among a set of possible metrics,
 \item[c)] Must be considered in relation to existing legal arrangements, which incorporate these social choices, and vary over time and space,
 \item[d)] May involve trade-offs between technical and legal constraints.
\end{itemize}

The article is intended as a contribution to the interdisciplinary constructs of fairness analysis in algorithmic decisions. It is clear that designing fair algorithms is not an easy task. Section 1 shows that technical choices in supervised learning have social implications that must be taken into account. Section 2 proposes a contextualized approach to the issue of unintended group discrimination, i.e., decision rules that are facially neutral but generate disparate impacts across social groups (as the case may be: gendered, racial or ethnic). The contextualization will focus on the legal systems of the United States on the one hand and Europe on the other. In particular, legislation and jurisprudence tend to promote different criteria of fairness on both sides of the Atlantic. Section 3 will be devoted to the explicability of algorithmic decisions; it will compare and attempt to cross-check legal concepts (in European and French law) with technical concepts and will highlight the plurality, and even the polysemy, of European and French legal texts concerning the explicability of algorithmic decisions. The conclusion will propose directions for future research.

\section{Fairness in machine learning: social implications of technical choices}

\subsection{Supervised learning}

Machine learning (ML) algorithms are increasingly used to make or assist decisions that affect people’s lives, in applications such as lending, pricing, hiring, criminal justice and medical diagnosis. In this paper, we focus on the main task of supervised learning, in which algorithms learn to predict an outcome variable $y$ from input variables $x$. These outcome variables can be quantitative (e.g., predicting house prices based on location and surface) or qualitative (e.g., predicting whether a loan applicant will repay or default, based on characteristics such as debt history and occupation). Supervised learning is the task of learning a prediction rule $h$, from a sample of labeled data $S = \{(x_i,y_i)\}_{i=1}^n$ called training dataset, so as to predict for each new $x$, an outcome $y$.

The relationship between inputs and outputs is characterized by an unknown function. The algorithm must choose $h$, from a given family of functions (the hypothesis class $\cal{H}$), which best approximates the unknown function on the training examples $S$. The main approach to choosing h is called empirical risk minimization (ERM). The algorithm is given a loss function $l(h(x),y)$ that quantifies how different the prediction $h(x)$ is from the true outcome $y$. ERM is the common practice of finding $h$ by minimizing the empirical risk $\hat{R}(h)$, i.e. the average loss over the training dataset $S$:

\[\min_{h\in\cal{H}}\Big\{\hat{R}(h):=\sum_{i=1}^{n}l(h(x_i),y_i)\Big\}\]

where $\cal{H}$ is the hypothesis class, i.e., the class of machine learning models considered, which can include more or less complex models such as decision trees or SVMs.

\subsection{Impact of underlying choices in supervised learning}

The common practice of ERM for supervised machine learning that we described above involves many choices that have an impact on the predictions and subsequent decisions:

\begin{enumerate}
 \item collecting and preparing the training dataset $S$,
 \item choosing a class of machine learning models $h\in\cal{H}$,
 \item minimizing an average risk $\hat{R}(h)$ (instead of e.g., the worst-case risk),
 \item choosing a loss function $l(h(x),y)$ (some losses are more robust to ``outliers'' than others),
 \item choosing an evaluation benchmark.
\end{enumerate}

In the following, we describe the different form of biases that may arise at each step and we classify them in two categories: data bias and algorithmic bias. In practice, ML pipelines are much more complex and bias is prone to arise at many other levels. We refer to \cite{Mitchelletal2021} and \cite{SureshGuttag2021} for a more complete taxonomy of sources of potential harms in machine learning-based decisions.

\begin{figure}
  \centering
  \includegraphics[scale=0.9]{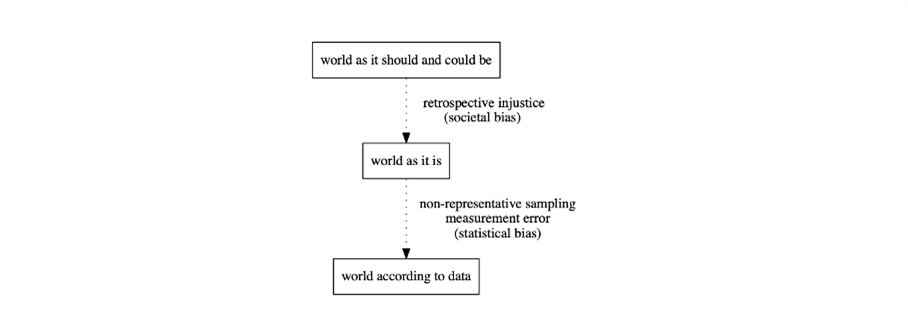}
  \caption{Illustration of data bias (from \cite{Mitchelletal2021})}
  \label{databias1}
\end{figure}

\begin{figure}
  \centering
  \includegraphics[scale=0.9]{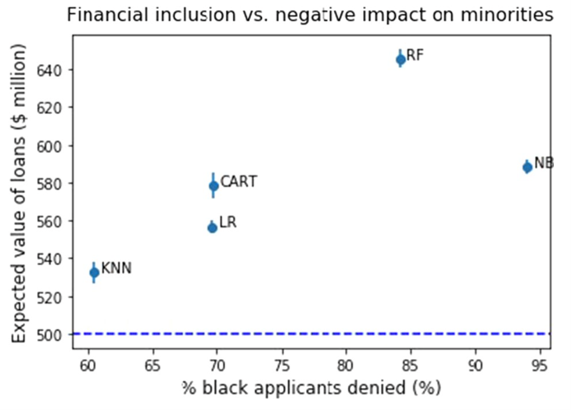}
  \caption{Financial inclusion vs. negative impact on minorities (from \cite{LeeFloridi2021})}
  \label{databias2}
\end{figure}

\emph{Data bias}. The presence of bias in ML-based decisions is commonly attributed to bias in the data. More precisely, bias arises in the process of generating the data that is used to train and evaluate ML models, and it may appear for a variety of reasons (see Figure \ref{databias1}). First, in statistics, sample or selection bias arises when the collected data-set under-represents some parts of the population. For example, most facial recognition technologies are trained on data-sets which are skewed towards light-skinned men (\cite{BuolamwiniGebru2018}), resulting in worse prediction performance for darker-skinned women. Similarly, object recognition data-sets are ``biased'' towards developed countries and fail to accurately detect object in lower-income households (\cite{deVriesetal2019}). Second, measurement bias arises when the labels y are only proxies of the true outcomes we want to predict. For example, when machine learning is used to predict crime in pre-trial detention problems, crime is in fact never measured and the algorithm is instead trained to predict arrests (\cite{CorbettGoel2018}, \cite{LumIsaac2016}). However, arrest rates strongly differ by neighborhoods and race, hence making arrests a biased proxy of actual crime. Third, the data may reflect societal or historical bias even if it is perfectly representative or measured. Societal bias occurs when real-world inequalities are structural, and when reproducing or exacerbating them undesirably harms a disadvantaged group. For example, even if we were able to perfectly collect data on the qualification of men compared to women on tech jobs (\cite{Dastin2018}), one should be careful of the potential harm or social acceptability of systematically predicting that men are more skilled than women on unseen cases. Since ML models are trained to fit the data distribution, they are prone to reflect bias in the data arising from sampling, measurement and historical processes (``bias in, bias out'', \cite{KallusZhou2018}).

Collecting more data is often seen as a straightforward solution to mitigating data bias, and bias in machine learning as a whole \cite{Hooker2021}. While being a relevant strategy to compensate for selection bias (\cite{Chenetal2018}), it cannot circumvent measurement bias and historical bias. To address those biases, authors proposed several techniques to learn fair representations of the data (\cite{Louizosetal2015}, \cite{Zemeletal2013}) or ``de-bias'' standard data representations (\cite{Bolukbasietal2016}, \cite{Zhangetal2018}).

\emph{Algorithmic bias}. Algorithmic choices, such as (2–5) in the list above, also impact the predicted outcomes of the algorithm, in a way that may affect different social groups differently. In the literature, when ML algorithms exacerbate existing bias in the data, this problem is referred to as bias amplification (\cite{Halletal2022}, \cite{Lloyd2018}, \cite{Zhaoetal2017}).

\begin{figure}
  \centering
  \includegraphics{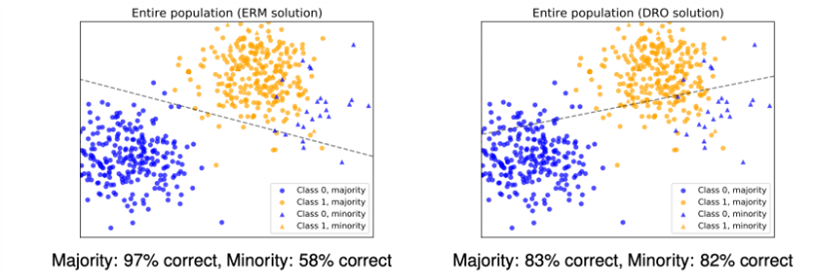}
  \caption{Illustration of ERM vs. DRO on synthetic data (from \cite{SlowikBottou2021})}
  \label{databias3}
\end{figure}

In this example, minimizing the average risk is harmful to the minority population, while minimizing the worst-case group risk equalizes error rates across groups.

First, the choice of machine learning model is known to impact the prediction behaviour, and in particular the quality of predictions on unseen data. Setting fairness considerations aside, a widely taught example is that varying the degree of a polynomial model leads to various levels of overfitting. As noted in \cite{Hooker2021}: ``our modeling choices [...] express a preference for final model behaviour.'' In the context of fair machine learning, the choice of ML model is an implicit choice for model behaviour on different social groups. Figure \ref{databias2} illustrates a bank loan example from \cite{Leeetal2020}, with different ML models leading to different trade-offs between the proportion of black applicants denied and the expected value of loans granted, showing that some models lead to high financial inclusion only at high cost for the disadvantaged group.

Second, the paradigm of ERM, which consists in minimizing averages, is itself a choice that impacts predictions. Alternative paradigms may have a different impact on disadvantaged groups. Such an alternative uses the framework of distributionally robust optimization (DRO) (\cite{BenTaletal2009}), and consists in minimizing the worst-case risk, rather than the average risk. For example, in Figure \ref{databias3}, \cite{SlowikBottou2021} show a synthetic example where using ERM yields large error on a minority group, while DRO equalizes error rates between the majority and minority groups, while maintaining a reasonably low overall error rate. Other authors advocated for similar “min-max” approaches to fairness (\cite{Dianaetal2021}, \cite{Hashimotoetal2018}, \cite{Levyetal2020}, \cite{Wangetal2020}).

Importantly, model evaluation is also a source of bias, leading to biased feedback loops that reinforce existing stereotypes. Standard benchmarks may misrepresent some sub-populations and single average performance metrics can hide poor performance on a minority group (\cite{SureshGuttag2021}). Existing audits for fairness evaluate models by quantifying disparities in performance across groups (\cite{BuolamwiniGebru2018}, \cite{Dattaetal2015}, \cite{Sweeney2013}), which can be seen as additional metrics to prevent against biased evaluation. Further, model cards were introduced as a standard for reporting multiple metrics for the evaluation of potential harm of models across different cultural and demographic groups, as well as their intersections (\cite{Mitchelletal2019}). These model cards have been increasingly adopted by ML practitioners (e.g., \cite{Menonetal2020}).

\subsection{Fairness in machine learning}

In the evidence of biases in ML predictions, researchers have been actively designing methods to mitigate them, giving rise to a large body of work on fairness in machine learning (which is extensively reviewed in \cite{Barocasetal2019}). Many fairness metrics have been proposed, and the most widely used criteria were shown to be incompatible (\cite{Chouldechova2017}, \cite{Kleinbergetal2016}), and to present risks for the populations they were meant to protect (\cite{CorbettGoel2018}). It is now acknowledged that mere technical approaches are superficial patches to address bias in machine learning (\cite{Selbstetal2019}), and that careful approaches to fair machine learning should take social and legal context into account.

\section{Problematizing the ethical question, between law and algorithmic technique: indirect discrimination}

The problem of equity is a classic one in economics and in moral and political philosophy. It is all the more complex in that it is closely linked to the question of social justice, which is the subject of several theoretical systems (utilitarianism, libertarianism, egalitarianism, marxism; \cite{ArnspergerVanParijs2003}). In the legal doctrine, Aristotle's formula of proportional equality states that ``things [and persons] that are alike should be treated alike, while things that are unalike should be treated unalike in proportion to their unalikeness'' is well known, even if legal doctrine and practice show its limits (\cite{Schieketal2007}, p. 27). Instead, the legal doctrine and practice retain other Aristotelian principles: corrective justice and distributive justice, whose distinction is the basis for the separation of private law (a contractual prejudice must be compensated - corrected) and public law (which organises rights of access to resources - hence the distributive aspect). Following on from these debates, the legal doctrine and practice argue that equality should be substantive rather than merely procedural, meaning that anti-discrimination law aims to transform social realities to bring about greater equality. However, ``Substantive equality is not a uniform concept. It comprises equality of results, equality of opportunities, equality in relation to substantive rights such as freedom of profession or capabilities, and equal respect ...'' (\cite{Schieketal2007} p. 28). In economics, legal science and moral philosophy, there is no single definition or criterion of fairness.
The field of machine learning also has a plurality of definitions and criteria of fairness\footnote{In this regard, Binns (\cite{Binns2018}) argues that ``fairness as used in the fair machine learning community is best understood as a placeholder term for a variety of normative egalitarian considerations'' which can be based on preference-based welfarism, Rawls' theory of justice or Sen's theory of capabilities.}.

This said, it should not be overlooked that research on fairness in machine learning is predominantly American or influenced by it; it takes as its background, mostly implicitly, the legal and institutional context of the US. Ethno-racial groups are statistically recognized in the US; there are extensive federal civil rights and non-discrimination laws for protected groups; the notion of disparate impact is a concept created by law. We question here the application of fairness approaches to other legal and institutional contexts, by developing the case of European anti-discrimination law (\ref{inddiscr}). The intersection between machine learning approaches on the one hand, and legal approaches on the other, is a necessity to create the conditions for algorithmic decision models consistent with anti-discrimination law. This implies defining the terms in which the relationship between law and algorithmic models can be posed, with the ultimate goal of algorithms that are consistent with the law and the values it carries (\cite{Koulu2021}) (2.2.).

\subsection{Governing indirect discrimination} \label{inddiscr}

We analyse the issue of fairness from two perspectives: a) indirect discrimination and disparate impact law by comparing the United States and the European Union (\ref{lawUSEU}); b) the gap between anti-discrimination law, whose implementation and interpretation by the courts is dynamic, and an ideal-typical machine learning formalisation (\ref{dynamic})

\subsubsection{Anti-discrimination law in the US and the EU: specific regimes} \label{lawUSEU}

Disparate impact is a common issue in technology and law. As mentioned above, the fact that the vast majority of machine learning research takes US anti-discrimination law as its background leads us to characterize the European legal context as well. We propose here, through a comparison of the American and European legal contexts, to: a) assess the relevance of work in ethical Machine Learning in the European context b) characterize ethical standards and measures of discrimination in the EU.

With regard to the United States, it should be noted that the notion of disparate impact emerged there, on the basis of the Civil Rights Act of 1964, Title VII of which specifies that practices which, under the guise of neutral rules, have a disproportionate impact on the protected class as compared to the unprotected class are prohibited - unless a legitimate interest (a business necessity) can be validly invoked.

In addition to Title VII, Title VI of the Civil Rights Act gives the federal government the right to sue for disparate impact. In addition, two laws from 1967 - the Age Discrimination in Employment Act and the Fair Housing Act - give individuals a right of action for disparate impact.

These provisions of federal law have been given concrete form by the implementation of a calculation rule to determine whether a situation of disparate impact can be observed in practice, known as the "80\% rule" (or 4/5 rule). Initially developed in the State of California, it was formalized in the State of California Guideline Selection Procedures in 1972. It was then codified at the federal level in the Uniform Guidelines on Employee Selection Procedures (1978) used by the US Equal Employment Opportunity Commission (EEOC), but also by the DoJ and the  DoL in their claims before the courts under Title VII. The 80\% rule is a ratio of the hiring rate of individuals from the unprotected class to individuals from the protected class: if a company hires 50\% of the male applicants for a job and 20\% of the female applicants, the hiring rate is equal to 50/20, i.e. 0.4: the hiring rate of women is 40\% of that of men. According to the EEOC guide, if the ratio of selection rates is less than 80\%, this indicates a discriminatory situation.\footnote{The relevance of the 80\% rule is debated, both legally and statistically (\cite{Peresie2009}). We will come back to this later.}

On the legislative side, the disparate impact doctrine was imported into the UK from the early 1970s, when the Griggs decision appealed to British political leaders during a visit to the US, to such an extent that they incorporated indirect discrimination into the first sex discrimination law (the Sex Discrimination Act of 1975) and into the revision of the Race Relations Act in 1976 (\cite{Suk2014}, p. 287). As will be discussed in more detail below, the enforcement of Title VII of the Civil Rights Act by the US courts has progressively introduced greater and greater restrictions on its invocation by victims of indirect discrimination: the plaintiff must establish precisely which of the companies' employment management rules is the cause of a disparate impact; provide compelling statistical evidence; and demonstrate that the employer refused to implement management rules that could have reduced the disparate impact. Even if the victim succeeds in overcoming these tests, the employer can show that the challenged rule or practice meets a business necessity.

The European Union has enacted a number of directives to combat discrimination, both direct and indirect, on the grounds of racial or ethnic origin, religion or belief, disability, age or sexual orientation: two directives on employment equality and race equality were adopted in 2000. In 2009, in the Lisbon Treaty, a horizontal clause was introduced to incorporate anti-discrimination into all EU policies and actions (article 10 of the TFEU\footnote{Treaty on the Functioning of the European Union}). The CJEU\footnote{Court of Justice of the European Union} has dealt with numerous discrimination cases, either in response to preliminary questions from national courts or on the merits.

Suk (\cite{Suk2014}) highlights how different US and EU anti-discrimination law is.
\begin{itemize}
  \item There is a significant body of European case law on equal pay for men and women, while the US federal courts have excluded it from the disparate impact system. In this respect, \cite{Selmi2006}, in an empirical analysis of litigation concerning disparate impact cases in the US, points out that it has had a clearly limited impact outside the context of written tests used in employee recruitment.
  \item The decisions of the CJEU are more protective of female employees than in the US, in particular because unlike the European Court, the US courts have refused to apply the disparate impact theory to part-time work situations (which are much more important for women than for men).
  \item	It is relatively easier for claimants to argue a prima facie case of discrimination in Europe than in the US: the US requirement to determine which specific provision, rule or practice would cause a disparate impact has no equivalent in Europe.
  \item Therefore, the CJEU focuses on equal pay for men and women; it is relatively silent on racial inequalities, whereas these are central in the US; in the US the disparate impact is applied to employment, excluding pay, i.e. to recruitment and promotion situations. Part-time work situations are not included in its scope (see table \ref{tableau 1}).
\end{itemize}

\begin{table}
 \scriptsize
  \centering
   \begin{tabular}{lll}
     \hline
     ~ & United States & E. U. \\ \hline
     Main focus & Racial inequalities & Equal pay for men and women \\
     ~ & Hiring and career & ~ \\
     ~ & progression & ~ \\ \hline
     Part-time work & Not taken into account & Taken into account \\ \hline
     Burden of proof (on the applicant)& Demanding and restrictive & Not very restrictive \\ \hline
     Justification for rules or practices that have a & Business necessity & Balanced approach to business \\
     disproportionate impact on employees & argument benefits & necessity in the case law of the \\
     ~ & employers & CJEU \\ \hline
     \hline
   \end{tabular}
  \caption{Comparison between United States and European Union.}\label{tableau 1}
\end{table}

\subsubsection{Disparate impact: a dynamic legal concept} \label{dynamic}

The question here is the correspondence between the (ideal-typical) computer approach and the judicial practice of disparate impact. The basic form of disparate impact is as follows:

\[ DI = \frac{(P(Y=1) | (S=0))}{(P(Y=1) | (S=1))}\]

It consists of comparing the probability of obtaining an outcome $(Y=1)$ according to whether the group is protected $(S=1)$ or unprotected $(S=0)$.
As envisaged above, the enforcement of these legal provisions has resulted in the application of a calculation rule to determine whether a situation of disparate impact has occurred (the "80\% rule").

The Uniform Guideline has been codified into the Code of federal regulation which article 1607 of Title 29 states that: ``A selection rate for any race, sex, or ethnic group which is less than four-fifths (or eighty percent) of the rate for the group with the highest rate will generally be regarded by the Federal enforcement agencies as evidence of adverse impact, while a greater than four-fifths rate will generally not be regarded by Federal enforcement agencies as evidence of adverse impact. Smaller differences in selection rate may nevertheless constitute adverse impact, where they are significant in both statistical and practical terms or where a user's actions have discouraged applicants disproportionately on grounds of race, sex, or ethnic group'' (29 CFR 1607).

In addition, the EEOC is mindful of the disparate impact that certain recruitment practices may have, such as criminal record and banking history (credit defaults), as African-Americans are over-represented in the prison population and in the defaulting debtor population. In the 4th Circuit Court of Appeals decision EEOC v. Freeman (2015), the judge struck down the EEOC's request to enjoin the company from conducting criminal and bank records searches. Judge Titus went so far as to write that the EEOC's application is a ``theory in search of its practice''. He added that employers face a tension between exposing themselves to potential liability if a blindly recruited employee is found to commit criminal or fraudulent acts and exposing themselves to prosecution by the EEOC.

The EEOC 80\% rule has limitations, which are recognized as such in the law. According to 29 CFR 1607, large differences in selection rates do not mean a disparate impact when they involve small numbers or are not statistically significant. In this case, the EEOC allows for statistical evidence over a longer period of time and/or to study the impact of a similar practice implemented in similar circumstances. The sensitivity to small numbers of employees makes the rule inequitable when it leads to very small firms being more exposed to liability for discriminatory impact than large firms with much larger numbers of applicants for employment and applicants being considered for employment (\cite{Peresie2009}\footnote{A numerical example can be used to establish this. Let us consider a large firm (LF) and a small firm (SF). The LF receives 20,000 applications, equally divided between men and women. The recruiter selects the CVs of 3200 women and 4000 men, i.e. 32\% and 40\%. The selection is then equal to 0.32/0.4 = 0.8, in accordance with the EEOC rule. The SF receives 20 applications, equally divided between men and women. The employer selects the CVs of 4 men (40\% of the applicants) and 3 women (30\% of the applicants), resulting in a selection ratio below the 80\% threshold: 0.3/0.4 = 0.75.}). Finally, the 80\% rule has its limits in terms of establishing causality, and the courts prefer to use statistical significance tests (\cite{Peresie2009}, p. 785).

In the common law system, such as the United States legal system, case law has a particular importance: being recognized as a source of law, it interprets statutes and gives them their practical meaning. The 1971 Supreme Court decision, Griggs v. Duke Power, was a significant advance in securing civil rights for African Americans. The firm in question conducted intelligence tests and required employees to have attained a high school diploma in order to be promoted to more lucrative positions. African-American employees were rarely promoted, while Caucasian employees were frequently promoted. In its decision, the Supreme Court ruled that if a practice that operates to exclude members of a protected group is not based exclusively on job performance, then it is prohibited. The promotion practice of the accused firm was not, in this case, based on sole work performance (\cite{Vinik2010}).

After Griggs, the US Supreme Court followed a more restrictive jurisprudence. In the second half of the 1980s, restrictions were placed on constitutional protections; then, in the 1990s, restrictions were introduced around the evidentiary requirements for plaintiffs. Finally, in the 2000s, case law has limited the scope of employers' actions to avoid racially disproportionate outcomes (such as in Ricci v. DeStefano) (\cite{Suk2014}).

The Supreme Court has barred disparate impact claims when they invoke the Equal Protection Clause of the 14th Amendment to the Constitution, holding that the clause covers only intentional discrimination (\cite{Suk2014}). In Arlington Heights v. Metropolitan Housing Corp. (1977), the Supreme Court held that ``Proof of racially discriminatory intent or purpose is required to show a violation of the Equal Protection Clause''. The same applies to the 5th amendment (``due process clause''): In Washington v. Davis (1976), the Supreme Court held that a disparate impact argument cannot be used in a Fifth Amendment claim unless the plaintiff demonstrates that racially neutral standards were used with discriminatory intent\footnote{Barocas \& Selbst (\cite{BarocasSelbst2016}) call such a practice ``masking'': ``... any form of discrimination that happens unintentionally can also be orchestrated intentionally'' (p. 692).}. A year later, in Pothard v. Rawlinson (1977), the Court held that Title VII of the Civil Rights Act does not make the physical requirements for entry into the career of a prison guard unlawful, even if they exclude 40\% of female applicants.

It was in the 1980s that case law took a restrictive turn, hardening the standards of proof imposed on plaintiffs. These restrictions culminated in Wards Cove Packing v. Atonis (1989) The Court has placed a very important restriction on disparate impact actions by establishing the evidentiary rule that the plaintiff must establish: (a) the employee must identify the specific practice or rule that caused the indirectly discriminatory impact, and (b) the employee must also prove that the employer refused to implement practices or rules that would have satisfied the plaintiff's grievances. In addition, the offending firm has the option of arguing in favor of the rule or practice causing a disproportionate impact, taking into account the ``necessity of business''. It should be noted that the requirement to precisely define the practice or rule causing a disparate impact has been extended to other areas, and has become a constraint on the ability to advance statistical evidence: in Texas Department of Housing \& Community Affairs v. Inclusive Community Project (2015), the Supreme Court held that a claim based on statistical parity must be dismissed if it does not establish which precisely defined public policy measure is alleged to cause a disproportionate impact. The Court also considered that housing policies must have room for flexibility, which is necessarily to achieve their legitimate interests (\cite{Vinik2010}).

A third phase of Supreme Court restrictions on disparate impact theory began in the 2000s. The most significant decision was Ricci v. De Stefano, issued in 2009 following the City of New Haven's (Connecticut) annulment of an internal competition for the promotion of the city's fire-fighters, as the success rate of white fire-fighters was twice that of African. Mayor John De Stefano considered that a disproportionate impact justified the cancellation of the exam which was challenged by the white fire-fighters and one Hispanic, including Ricci, who had passed the tests and should have been promoted. The Supreme Court ruled that the decision to overturn the test violated Title VII of the Civil Rights Act, as the City of New Haven did not have a ``strong basis of evidence'' indicating that it would have faced disparate impact liability if it had promoted white and Hispanic fire-fighters over African-American fire-fighters. In sum, the Supreme Court has limited the remedies available to employers to avoid racially disproportionate outcomes.
Thus, while the terms of the law do not give a fixed content to the disparate impact doctrine and its empirical scope is shaped (narrowly) by the courts, its modelling in computer science is based on an ideal-typical formulation that is insensitive to its implementation in the real world.

\subsection{Bringing anti-discrimination law and algorithms into dialogue: how? For which purposes?} \label{bringing}

An increasing number of authors are endeavoring to link legal and technical concepts related to issues of fairness and direct and indirect discrimination. Public policy issues are raised, which call for a bridging of disciplines. The fact that there is a growing body of literature in this field is good news. However, it seems to us that the issues and objectives targeted by the authors are diverse and reveal different postures that it seems important to identify. We identify four types.

\subsubsection{Comparison of concepts per se}

Xiang and Raji (\cite{XiangRaji2019}) compare the concepts of machine learning and US anti-discrimination law on several dimensions. We present here their comparison with comments.

\begin{itemize}
  \item \textbf{Procedural fairness:} In machine learning, it refers to the identification of the characteristics of the inputs (in particular the proxies used) that lead to a particular outcome of the model, the focus being on the algorithm itself and its predictions. In contrast, for law, it is about the governance principles of the decision-making process. The focus is on the "system surrounding the algorithm and its uses" rather than on the algorithm itself.
  \item \textbf{Discrimination:} In machine learning, it is often presented as the product of an unfair correlation between protected class variables and a metric of interest such as outcomes, false positive rates between classes, or the like. For the law, it is a social issue for which rules are produced with the aim of suppressing discriminatory intentions or causality. Note that Xiang and Raji do not mention indirect discrimination, which does not involve discriminatory intent.
  \item \textbf{Protected Class/Sensitive Attribute:} While this is a major issue for law, machine learning research makes it a matter of features that should not be involved in algorithmic decision making, without the need to take law into account.
  \item \textbf{Anti-classification and anti-subordination:} The notion of anti-classifi\-ca\-tion in law simply means that the state cannot classify individuals according to their race, gender, age, etc. It is consistent with the concept of fairness through unawareness with which the machine learning community working on fairness is familiar. The principle of anti-subordination refers to equality of rights as an objective, which is however unattainable in a society with strong social stratification unless the law aims to strengthen the position of the less favored groups. According to \cite{XiangRaji2019}, this is a dimension rarely considered in the machine learning literature. There are indeed minority approaches, such as "min-max" approaches, which aim to minimize misclassification for the most disadvantaged social groups (see the DRO-type approaches mentioned earlier). \cite{Doetal2021}, using Lorenz dominance, consider fairness in the sense of improving the utility of the worse-off, following the transfer principle in economics.
  \item \textbf{Affirmative action:} Machine learning research is open on this matter, in its developments on demographic parity; in law, it is subject to varying interpretations over time.
  \item \textbf{Disparate treatment and disparate impact:} From a legal point of view, disparate treatment refers to intentionally discriminatory treatment; for machine learning, intention is not a relevant dimension, since it arises when a protected attribute is used in the decision process, and its use can be avoided. In law, \cite{XiangRaji2019} (wrongly) consider disparate impact to be illegal if it is intentional, whereas technically disparate impact arises when outcomes between subgroups differ, even without intentionality.
\end{itemize}

Ultimately, from the perspective of interdisciplinarity, it is very important to understand how the same concept is shaped in both fields and what meaning is attached to it. However, this is not an end in itself. It is a necessary task, but it should only be a prerequisite for a reciprocal opening up of the fields, of which we will see some possibilities below.

\subsubsection{Qualifying legal norms based on technical indicators: measures of group discrimination and anti-discrimination law}

In \cite{Pedreschietal2012} a legally based classification of indicators for measuring discrimination is suggested. They start from the idea that the interpretation of legislation leads to different measures of discrimination and different rankings of possibly discriminatory contexts. The authors use a contingency table to characterize four situations, depending on whether the group is protected or not, whether the decision is positive or negative, and to define a series of discrimination indicators.

\begin{table}
 \small
  \centering
   \begin{tabular}{l|ll|l}
     ~ & \multicolumn{2}{|c|}{benefit} & ~ \\ \hline
     group & denied & granted & ~ \\ \hline
     protected & $a$ & $b$ & $n_1$ \\ \hline
     Unprotected & $c$ & $d$ & $n_2$ \\ \hline
     ~ & $m_1$ & $m_2$ & $n$ \\ \hline \hline
     \multicolumn{4}{c}{$p1 = \frac{a}{n_1}\;\;\;\;p2 = \frac{a}{n_2}\;\;\;\;                p =\frac{m_1}{n}$} \\
     \multicolumn{4}{c}{$RD = p1 - p2\;\;\;\;RR = \frac{p1}{p2}\;\;\;\;RC = \frac{1-p1}{1-p2}\;\;\;\;OR = \frac{RR}{RC} = \frac{a/b)}{c/d)}$} \\
     \multicolumn{4}{c}{$ED = p1-p\;\;\;\;ER = \frac{p1}{p}\;\;\;\;EC = \frac{1-p1}{1-p}$} \\ \hline
   \end{tabular}
  \caption{Discrimination measures (source: \cite{Pedreschietal2012}) \\
$P1$: proportion of benefit denied for the protected  group \\
$P2$: proportion of benefit denied for the unprotected  group \\
$P$: proportion of benefit denied for the whole subset \\
$RD$ (risk difference) is the absolute risk difference: $RD = P1-P2$ \\
$RR$ (relative risk) is the relative risk: $RR = \frac{P1}{P2}$ \\
$RC$ (relative chance, or selection rate) is the chance to obtain a positive decision: $RC = \frac{(1-P1)}{(1-P2)}$ \\
$OR$ is the odds ratio: $OR =  \frac{p1(1-p2)}{p2(1-p1)}$ \\
$ED$ (extended difference): $ED = P1-P$ \\
$ER$ (extended ratio): $ER = \frac{P1}{P}$ \\
$EC$ (extended chance): $EC = \frac{1-P1}{1-P}$}
\label{discrimination}
\end{table}

\cite{Pedreschietal2012} argue that  ``From a legal point of view, several measures are adopted worldwide. UK law mentions risk difference, EU Court of Justice has given more emphasis on the risk ratio, and US laws and courts mainly refer to the selection rate'' (ibidem, p. 2). But in the conclusion of their article the diagnosis is different: the European legal system is connected to the RC (relative chance) and the American system to the RR (risk ratio), which casts serious doubt on the coherence of their analysis of legal systems.

It is indeed a bit adventurous to univocally associate a discrimination measure with a national or regional legal system. For example, the EU Equality Directives of the 2000s are predominantly concerned with differential risk (to establish prima facie discrimination it is sufficient to establish that a provision, criterion or practice places people of racial or ethnic origin at a disadvantage compared to others - \cite{Schieketal2007}, $\S$ 13.072). Instead, the case law of the CJEU refers to the principle of proof of a disproportionately disadvantageous position of a group in order to qualify a situation as indirect discrimination, which refers to a relative risk (\cite{Schieketal2007} $\S$ 12.125 - gender pay inequalities and $\S$ 12.659 - more general doctrine). Reference can be made to the Kirshammer-Hack decision of the CJEU (30 November 1993): in this case, a part-time employee in a company with less than 5 employees was dismissed without compensation, as German law only guarantees the right to compensation to companies with more than 5 employees; the dismissed employee considered that she had been discriminated against in view of the fact that more women than men work in very small companies and on a part-time basis. The CJEU held that there would be discrimination between men and women if it were established that small businesses employ a significantly higher percentage of men than women, which was not statistically established by the claimant.
In the same vein, the CJEU decision Krüger (9 Sept. 1999), was given in response to a preliminary question from the Munich Labour Court: is the exclusion of part-time employees (provided for in the collective agreement) from the benefit of a wage bonus discriminatory? The CJEU held that if the national court finds that the rule, even if formally gender-neutral, actually affects a considerably higher percentage of women than men, then it is sex discrimination.

There is a large body of work on the metrics of discrimination, both direct and indirect (\cite{AbuElyounes2020}; \cite{Besseetal2018}; \cite{Chouldechova2017}; \cite{Dwork2012}). We confine ourselves to indirect discrimination, for which several fairness criteria have been formalised, and which are incompatible with each other (Table \ref{parity}).

\begin{table}
 \footnotesize
  \centering
   \begin{tabular}{|l|l|}
     \hline
     Statistical parity & $\forall a,a'\;P(\hat{y}=1|A=a)= P(\hat{y}=1|A=a')$ \\
     Equality of opportunity & $\forall a,a'\;P(\hat{y}=1|y=1,A=a)=(\hat{y}=1|y=1,A=a')$ \\
     Equalized odds & $\forall i\in\{0,1\},a,a'\;P(\hat{y}=1|y=i,A=a)= (\hat{y}=1|y=i,A=a')$ \\
     Group calibration & $\forall s\in\mathds{R},a,a'\; P(y=1|S=s,A=a)=P(y=1|S=s,A=a')$ \\
     \multicolumn{2}{|l|}{$\hat{y}$ classifier response; $y$ target value of the original data; $A$ protected attribute;} \\ \hline
   \end{tabular}
  \caption{Main indirect discrimination metrics (adapted from \cite{Wachteretal2021})}\label{parity}
\end{table}
	
Abu Elyounes (\cite{AbuElyounes2020}) considers that, in law, group fairness aims to improve the position of disadvantaged groups and to achieve substantive equality between them, without being limited to mere procedural or formal equality. Indeed, disparate impact arises in the presence of a formally non-discriminatory rule or practice based on the procedural principle of equal treatment of persons in similar situations. Starting from the idea that fairness is contextual,
Abu Elyounes' approach is to determine the conditions under which indirect discrimination metrics have relevance for legal approaches (table \ref{groupfairness}).

A concrete example of the issues involved in choosing one metric among others is the controversy over the COMPAS criminal recidivism prediction application, which has been criticized by ProPublica as racially discriminatory. For example, the algorithm misclassifies African-Americans as future criminals at twice the rate of whites; whites are misclassified as low recidivists (\cite{Angwinetal2016}; \cite{Larsonetal2016}); other authors (\cite{Chouldechova2017}; \cite{Rahman2020}) believe that the COMPAS model is calibrated, showing that the recidivism risks of African-Americans and Caucasians are equal across all recidivism risk scores. In this respect, \cite{AbuElyounes2020} is correct in considering that Propublica retains the equal opportunity criterion, while COMPAS argues that its algorithm is fair because it is calibrated.

\begin{table}
 \scriptsize
  \centering
   \begin{tabular}{l|l|l}
     \hline
     Sub-notion & Corresponding Legal Mechanism & Example of implementable case \\ \hline
     Decoupling & Affirmative action (as separate & When the minority group is very small and has \\
     ~ & but equal) & unique characteristics like women in the criminal \\
     ~ & ~ & justice system \\ \hline
     Statistical or & Affirmative action (preferably & Cases where affirmative action was approved by law \\
     conditional parity & through critical diversity) & like hiring and school admission \\ \hline
     Equal & Affirmative action (via equality of & When fixing on the outcome is sufficient and does \\
     opportunity & opportunity) & not require fixing the process that led to this \\
     ~ & ~ & outcome \\ \hline
     Equalized odds & Achieving equality by equalizing & When it is possible to achieve the right balance \\
     ~ & the false positive and false & between the two types of errors \\
     ~ & negative errors & ~ \\ \hline
     Calibration & Achieving equality by statistical & High stake cases that society is willing to give up on \\
     ~ & significance & equalizing the error rates \\ \hline
     Multi-calibration & Achieving equality by statistical & Pretrial, but it should be applied cautiously since it is \\
     ~ & significance, and accounting for & a new notion \\
     ~ & intersectoriality & ~ \\ \hline
   \end{tabular}
  \caption{Group Fairness Concepts and corresponding legal mechanisms (extracted from \cite{AbuElyounes2020}}\label{groupfairness}
\end{table}

\subsubsection{Algorithms that comply with the law?}

\cite{Wachteretal2020} find that there is a gap, or even incompatibility, between European legal notions of discrimination and existing work on algorithmic fairness and make proposals to reduce this gap.
As far as anti-discrimination law in the EU is concerned, the Equality Directives are drafted with a high level of generality; the Member States must transpose them into national law, and they do so with a very variable level of generality or precision (in particular as regards the conditions for statistical proof); the rulings of the CJEU are not constant over time as regards the conditions for proving indirect discrimination, even though a ``gold standard'' can be found in the European case law. Legal regulation therefore provides flexibility (via the transposition of directives and the case law of the CJEU, which is often marked by intuitive reasoning rather than a defined and stable metric)\footnote{CJUE 19 April 2012, case. C 415-10, Galina Meister v. Speech Design Carrier Systems GmbH: ``indirect discrimination may be established by any means, and not only on the basis of statistical evidence''.} whereas research on algorithmic fairness is based on a search for precision and consistency. On the one hand, ``European conceptualization of discrimination ... is contextual'' (\cite{Wachteretal2018}) while on the other hand automatic methods to detect and correct discriminatory decisions need clear rules.

The aim of \cite{Wachteretal2020} is to clarify how to construct considerations of fairness in the automatic decision, which respects as much as possible the contextual approach of European law, in particular of the CJEU. However, it is possible to find a 'gold standard' in European law. This gold standard was laid down in the Seymour-Smith decision of the CJEU, which is that the comparison between the discriminated and the non-discriminated group is the best method. According to \cite{Wachteretal2020}, the only fairness metric compatible with this principle is (conditional) demographic (dis)parity, formalized as follows (for a given attribute):

\[A=\frac{\mbox{No. of protected people in the advantaged group}}{\mbox{Total No. of people in the advantaged group}}\]

\[D =\frac{\mbox{No.of protected people in the disadvantaged group}}{\mbox{Total No.of people in the disadvantaged group}}\]

As soon as $D > A$, there is a demographic disparity. \cite{Wachteretal2020} add another test, called "Negative dominance", which occurs if $D > 50\% > A$: this test does not exist in the literature on algorithmic fairness. The test is done in two steps: a) the majority of the disadvantaged group must not belong to the protected class, b) only a minority of the protected class belongs to the disadvantaged group.

In another contribution, \cite{Wachteretal2021} elaborate on the role that anti-discrimination law can play in helping to define a metric that is consistent with its aims of promoting substantive rather than procedural and formal equality. They examine the compatibility between the fairness metrics used in machine learning and the aims of European anti-discrimination law. The latter aims, beyond the prevention of discrimination, to change society, public policies and practices to ``level the playing field'' and achieve substantive equality. While machine learning is about designing fairness techniques that solve problems of bias in the training data, knowing that these data reflect historical social stratifications, these techniques can be described as ``bias preserving''; conversely, the purpose of law is to transform these historical societal biases (``bias transforming'')\footnote{``Bias preserving'' fairness metrics seek to reproduce historic performance in the outputs of the target model with equivalent error rates for each group as reflected in the training data (or status quo). In contrast, ``bias transforming'' metrics do not blindly accept social bias as given or neutral starting point that should be preserved, but instead require people to make an explicit decision as to which biases the system should exhibit (\cite{Wachteretal2021}).}. Technically, equalized odds, equal opportunity and calibration, and others, fall under an orientation ``bias preserving'' (\cite{Wachteretal2021}). They estimate that out of the 20 fairness metrics existing in the literature, 13 of them are bias preserving in the sense that they are satisfied by ``matching error rates between groups'' while the 7 metrics\footnote{precisely: statistical parity, conditional statistical parity, fairness through unawareness, fairness through awareness, counterfactual fairness, no unresolved discrimination, no proxy discrimination, path base causal reasoning.} ``bias transforming'' are satisfied by ``matching decision rates between groups''. As bias preserving metrics carry the risk of consolidating situations of social injustice and discrimination, thus maintaining the status quo, they are not in line with the core of European anti-discrimination law, which aims at achieving substantive equality (ibidem). \cite{Wachteretal2020}, \cite{Wachteretal2021} recommend the use of Conditional Demographic Disparity (CDD) which they consider to be compatible with the aims of European law and are therefore ``bias transforming'': \emph{``CDD treats all people (groups) as equal, meaning they should be treated the same. The test flags up any disparity between groups that remains once an appropriate conditioning variable has been applied. This notion of fairness follows the Aristotelian postulate of treating ‘like cases alike’ and enables formal equality. At the same time, CDD enables substantive equality by flagging up … any relative disparity between groups in a given population over a set of decisions or other outcomes''} (\cite{Wachteretal2021}).

\cite{Xiang2021} makes proposals on how to reconcile technical and legal approaches to algorithmic banking, based on the observation that the proliferation of automatic decision models poses a serious risk of increasing social inequalities. She discusses technical approaches that could comply with US anti-discrimination law and reduce the risk that algorithms that exacerbate inequality will be found to comply with the law. Xiang argues for causal inference, consistent with Texas Dpt of Housing and Community Affairs v. Inclusive Communities Project, Inc, which posits a 'causal connection' between the decision-making process and the existence of a disparate impact. For her, causal inference implies the use of a counterfactual, since it involves ``comparing what happened in the real world with what would have happened in a counterfactual world with different conditions'' (\cite{Xiang2021}).

\subsubsection{The law in confrontation with algorithmic discrimination}

Algorithmic discrimination raises questions about the capability of victims to go to court to challenge the differential treatment to which they are subjected. Access to the characteristics of the algorithmic treatment, or even knowledge of the existence of an automatic decision-making process, is in practice not ensured for victims. Hacker (\cite{Hacker2018}) raises the problem of the limited effectiveness of anti-discrimination law in Europe in the presence of algorithmic discrimination. According to him, the problem can be solved by unbundling anti-discrimination law and personal data protection as provided for in the GDPR; such an unbundling is conducive to a fruitful cooperation between lawyers and computer scientists likely to lead to the emergence of algorithms ensuring an ``equal protection by design'' (\cite{Hacker2018}, p. 25). The mobilization of the GDPR can move the cursor upstream of litigation, i.e. to a certain extent to encourage the design of decision rules that prevent situations of discrimination.  Thus, the lever of personal data protection can have beneficial effects in terms of anti-discrimination, without the need to strengthen data protection law. The expected beneficial effects can be achieved with the law as it is, i.e. without the need to amend the GDPR (\cite{Hacker2018}, p. 25).

The GDPR contains recitals that set out key principles, including $n^o$39 which states that all processing of personal data should be lawful and fair; and especially $n^o$71 which states that ``the controller should use \textbf{appropriate mathematical or statistical procedures} for the profiling, implement \textbf{technical and organisational measures} appropriate to ensure, in particular, that factors which result in inaccuracies in personal data are corrected and the risk of errors is minimised, and that prevents, inter alia, \textbf{discriminatory effects} ...'' (we underline).

The GDPR contains several instruments that can bring about these beneficial effects: the individual right of access to data; the data protection impact assessments (DPIAs) and audits that were proposed by Article 29 of the Working Party\footnote{The Article 29 Working Party is an independent European working party that dealt with issues relating to the protection of privacy and personal data until 25 May 2018 (entry into application of the GDPR)} Guidelines on Automated individual decision-making and Profiling. The guidelines proposed, among other things: an ``algorithmic auditing'', consisting of ``testing the algorithms used and developed by machine learning systems to prove that they are actually performing as intended, and not producing discriminatory, erroneous or unjustified results''; These proposals have been incorporated into Recital 71 and Article 22 of the GDPR.

Hacker believes that the GDPR incorporates key principles - such as data must be lawful, fair, accurate and the avoidance of discriminatory effects - which are an important basis for algorithmic fairness. Other elements of the GDPR are the provisions of Article 15 (Right of access by the data subject) and Article 22 (Automated individual decision-making) on substance; but Hacker considers that these provisions, in particular Article 15(1)(h), should be supplemented by a ``public enforcement that aims at uncovering the right metrics and exact causes of discrimination'' (\cite{Hacker2018}, p. 27). Public enforcement refers to Articles 83 (General conditions for imposing administrative fines) and 58 (powers of supervisory authority) as well as Article 35 (Data protection impact assessment). Thus, ``National data protection authorities should moreover make use of algorithmic audits and data protection impact assessments, according to Article 58(1)(b) and Article 35 and seq. GDPR, to uncover the causes of bias and to enforce adequate metrics of algorithmic fairness'' (\cite{Hacker2018}, p. 35).

The following developments focus on explicability, comparing the legal and machine learning models, around the idea of fairness through explainability.

\section{Explainabillity of algorithmic decisions: necessity of a techno-legal clarification.}

Applied in the field of algorithms, the notion of explainability is a legal and technical notion still under construction. This notion is at the heart of an important field of AI known as Explainable AI (XAI), initiated forty years ago\footnote{The emergence of this expression is often refered to two articles: \cite{Scottetal1977} and \cite{Xuetal2019}} but it becomes more important with the development of algorithmic decisions.

In everyday language, the term explainability refers to the ability to explain, to make intelligible or understandable. From the outset, this term has several dimensions. It is a question of explaining the why and/or the how of something. The result of this explainability will be to provide explanations. To speak of explainability necessarily implies delimiting the object, taking into account the addressees, and determining who will have to provide these explanations, and even in what form. European and French law insist on the fact that algorithmic decisions must be explainable, but without giving a clear definition or specifying its modalities. In AI, explicability is differentiated from interpretability, which introduces confusion in the understanding of the meaning of notions in the various domains.

The purpose of this part is to draw up a cross-inventory of the state of the law and of the technique with regard to explainability. One of the first sources of difficulty in conducting such a dialogue lies in the need to agree on a common vocabulary, knowing moreover that within each discipline, there is no unanimity on the meaning of such or such terminology. Therefore, it seems necessary to confront the legal approach with the technical one by drawing up an inventory of the state of research in AI on explainability. The objective is to contribute to a better understanding of how explainability could prevent and counter bias, and thus be a means of ensuring the fairness of algorithms.

\subsection{Explainability: a notion used but not defined by the legislator}

To our knowledge, there is no general and abstract definition of the notion of explainability by the legislator, whether in European or French law. Moreover, this term has only recently been explicitly enshrined in French law in Article 17 of the Bioethics Act of August 2, 2021. Neither the Digital Republic Act (Loi sur la République Numérique - LRN), nor the Data Protection Act (Loi Informatique et Liberté – LIL) include such a reference\footnote{Yet, in French law, this term is commonly used in particular in connection with articles L-311-3-1 and R311-3-1-1 of the Code of Relations Between the Public and the Administration as well as article 47 of the Loi Informatique et Liberté introduced to implement article 22 of the GDPR.}.

Nonetheless, in European and American law, there are legal obligations of explainability essentially linked to algorithmic decisions. In the United States, for example, a constitutional right to explainability has been enshrined in case law in two particular cases. The first case concerned an algorithm for ranking teachers. The judge considered that ``without access to value-added equations, computer source codes, decision rules, and assumptions, teachers could not exercise their constitutionally-protected rights to due process''\footnote{Local 2415 v. Houston Independent School District, 251 F. Supp. 3d 1168 (S.D. Tex. 2017), p. 17.}. The second case involves the COMPAS predictive justice algorithm. If the Wisconsin Supreme court\footnote{Cf. State of Wisconsin v. Loomis, Supreme Court of Wisconsin, n° 2015AP157-CR, July 13, 2016.} did not require the release of the source code of the COMPAS algorithm it enshrined an overall explainability requirement to allow judges to better assess the accuracy and weight to be given to the risk score in sentencing (see \cite{Beaudouinetal2020a} and \cite{Beaudouinetal2020b}). In the banking sector, there is also an obligation of explainability for the granting of loans. The bank must give specific reasons for the refusal of loans\footnote{Fair Credit Reporting Act (2018).}. Finally, some states have also introduced explainability requirements, such as Washington State's facial recognition law\footnote{See \url{https://www.dlapiper.com/en/us/insights/publications/2020/04/in-washington-
states-landmark-facial-recognition-law-public-\\
sector-practices-come-under-scrutiny/}}.

In the EU, there are also several explainability requirements, the most well-known of which was introduced in Article 22 of the GDPR\footnote{On this provision and the differences in wording with the article in Convention 109 of the Council of Europe, see \cite{Tambou2020}, p. 209 and following.}. It is important to note that the scope of this obligation is narrowed in two ways: first, to systems that use personal data - non-personal data is not governed by the GDPR; second, to automated individual decisions, including profiling (see table 6, APPENDIx)..

\subsection{Crossed views on notions: decision, profiling, explainability, interpretability}

What deserve attention are the terms in which law, especially European law (with the GDPR and the upcoming AI Regulation) represents the object of explainability standards: the decision. This representation is not naturally consistent with approaches to decision and decision making in computer science. It is the same with the term ``explainability''. It is therefore useful to compare the notions in law and in computer science.

\subsubsection{Decision and decision process}

In computer science, a decision is the response of an algorithm (implemented as software) to a query. From the point of view of a mathematical formalism, a decision is the partition of a given set. A partition consists in the separation of the set into subsets with an empty intersection and whose union constitutes the set. Each subset is called an equivalence class. If these classes are ordered and defined with reference to a norm we call the decision a rating. If the classes are ordered but not defined with reference to a standard we call the decision a ranking. If the classes are not ordered, but are defined in reference to a standard we call the decision an assignment; finally, if the classes are not ordered and are not defined in reference to a standard we call the decision a clustering.

The activities of an entity (human or artificial) that lead to a decision are a decision process. From this point of view the activity of an algorithm that has to calculate for a set of objects a score (of something) is a decision process. In fine, the algorithm makes a decision (e.g. the outcome of these calculations).

However, in most cases, this activity is part of a larger process that consists of assisting another entity that, in turn, must take a decision. We call this set of activities a decision support process. The decisions made in a decision support process are ``recommendations''. It may be possible for decision support processes to be used in another entity's decision process. It is possible to go as far as an intertwining of decision processes and decision support processes that produce a cascade of recommendations ... up to a decision that we will call ``final'' ... The notion of liability generally applies to this ``final decision'' ... even if in reality any decision (mathematically defined) can be considered as carrying liabilities. For example, when a bank employee receives a credit application, an algorithm calculates the applicant's credit score. In such a decision process, the algorithm ``decides'' the credit score (decision) and transmits it to the employee (recommendation) who in turn decides to grant or refuse the credit (final decision). It is essentially to this ``final decision'' that the law relating to the explainability of algorithms focuses.

\subsubsection{Profiling}

From a computer science viewpoint, profiling is a clustering decision problem where individuals characterized by a set of ``external'' attributes (such as demographics) and by a set of behavioral attributes (such as buying preferences) are grouped in classes of ``similarity''. In case the classes of potential behavior are predefined we turn having an ``assignment'' problem. In the first case we do not know a-priori which are the equivalence classes which define the correspondence between external and behavioral attributes and the clustering allows to ``discover'' them. In the second case we have a hypothesis about how many behavioral patterns are possible among the observed population and we test it. The result of a profiling exercise is that all individuals clustered or assigned to the same profile (behavioral pattern) are supposed to behave similarly and under such a perspective can be targets of actions aiming at influencing or modifying this behavioral pattern.

The notion of automated individual decision is not been defined in the GDPR, whereas the notion of profiling is defined as a form of automated processing aimed at evaluating a natural person, whether to analyze or predict his or her work performance, economic situation, behavior, etc.\footnote{See GPDR, article 4.}. Therefore, the legal notion of automated decision does not overlap with the distinction made in AI between ``automated'', ``autonomous'' and ``algorithmic'' decision-making (\cite{BrkanBonnet2020}). In AI, the notion of an automated decision means that the decision has been made on the basis of a series of precise predefined actions without further intervention by a human. These decisions would be easily predictable. The notion of autonomous decision implies that only the general objectives have been established by a human, leaving it to the machine to determine how to achieve them. Yet, although referring to the term automated decision, the Article 22 of the GDPR is about algorithmic decisions (from the AI perspective). This more general notion simply means that a decision has been made with the help of an algorithm. It is therefore a notion that encompasses both autonomous and automated decisions.

It should be emphasized that not every algorithmic individual decision is covered by Article 22: only those that are exclusively based on automated processing and that produce legal effects concerning the person or significantly affect him or her in a similar way are concerned. This last expression is a matter of debate among legal scholars. Indeed, the prohibition of exclusively automated processing only concerns decisions that have a significant impact. This goes without saying when the decision has legal effects such as the cancellation of a contract, the right or refusal of a social benefit granted by law such as a family allowance, or a housing allowance, the refusal of entry into a country ,etc. On the other hand, it is not always easy to determine what is an automated decision that, without having a legal effect, affects the data subject ``similarly significantly affects him or her''.

Recital 71 of the GDPR simply states ``automatic refusal of an online credit application or e-recruiting practices without any human intervention''\footnote{On the difficulties of interpreting this notion of significantly affecting in a similar way to a legal effect see the Guidelines on Automated individual decision-making and Profiling for the purposes of Regulation 2016/679, revised version of February 6, 2018, p. 23.}.  This covers the loss of a chance, of an opportunity. This concerns decisions concerning access to a service in the field of health or education.

\subsubsection{Explainability  and interpretability}

For some authors, ``explainability'' and ``interpretability'' are interchangeable notions (\cite{Beaudouinetal2020a}, p. 8), while others make a distinction. The common point between these two labels is that in both cases the aim is to make the decisions made by an algorithm understandable. Nevertheless, in the field of AI, some authors consider that interpretability aims at globally evaluating the process of a decision, i.e., in reality, at making the model used understandable. In other words, interpretability answers the question of how an algorithm makes a decision in a general way.
In other terms we can distinguish three levels which can technically be considered as explanations for a given algorithm and its execution.

\begin{enumerate}
  \item The first one is \textbf{descriptive}: it consists in reconstructing and tracing each single step the algorithm performs from receiving an input until it delivers the foreseen output (computer scientists collect this information in what they call a .log file). It is essentially a descriptive explanation of what the algorithm did or does and most of the times is interpretable only by coding specialists.
  \item The second one is \textbf{logical}: it consists in reconstructing the reasons for which an algorithm performs a given step, while it runs. It is essentially a logical explanation through which each single step can be viewed as a causal relation. Such explanations are useful in order to show that the output of the algorithm is logically related to the input. Formal verification of algorithms and security checking use this type of explanation in order to demonstrate that a given algorithm actually does what the specifications were expecting to happen, and that the execution is robust, secure, trustworthy, etc.
  \item The third one is \textbf{argumentative}: it consists in providing the ultimate reasons for which an algorithm computes a certain output, given a certain input. Such reasons are essentially of two types: the data provided, and the procedure used. A typical example is the result of a voting procedure: the result depends on the ballots casted by the voters, but also on how the majority is computed with that particular procedure. An argued explanation always ought to provide both of them. This type of explanation becomes problematic when the algorithm modifies the execution each time depending on knowledge which cumulates from previous executions or when the algorithm includes components for which is practically impossible to make a logical connection between input and output (black boxes).
\end{enumerate}

For \cite{Besseetal2018}, a decision rule is interpretable if we understand how it associates a response with observations (e.g., a decision tree); it is explicable if we understand on what elements the decision is based on. According to \cite{Stoicaetal2017}, interpretability ``means that the output of the AI algorithm is understandable to a subject matter expert in terms of concepts from the domain from which the data are drawn'' whereas explainability ``means that one can identify the properties of the input to the AI algorithm that are responsible for the particular output and can answer counterfactual or ‘what-if’ questions''.

In contrast, explicability goes further because it involves specifying in a concrete case which specific variables were decisive in making a particular decision. Substituting the semantic distinction between interpretability and explicability, some authors in machine learning (\cite{Guidottietal2018}) oppose global explicability to local explicability. Global explainability aims at explaining the whole algorithm, while local explainability is the ability to explain a specific algorithmic decision. We can also distinguish between ``post-hoc explanations'' and ``build-interpretable models'' (\cite{Barale2021}).

Post-hoc  explanations are constructed through statistical analysis after the data have been revealed (and the hypothesis tested). They can positively be used in order to deep the understanding of a given result but can also negatively contribute to constructing meaningless correlations. Build-interpretable models are models constructed in such a way that explanations are computed (or collected) while used. This might improve efficiency in computing explanations, but present two weaknesses: it may produce out-of-the-context meaningless interpretations and it can conceal other possible explanations for which no provision has been considered within the specifications.

These semantic variations are not really found in legal doctrine. Most institutional documents and the legal literature use the term “explicability”, sometimes the word ``intelligibility'' (\cite{CNIL2017}, p. 53). However, it is acknowledged that there are also two levels of explicability requirements. On the one hand, there is an ex ante explicability, which aims to inform the individual about the logic of the algorithm and which is closely linked to a right to information. On the other hand, there is an ex post explicability, which aims at explaining to an individual why an algorithm has made a specific decision. The debate in the legal sphere is to what extent the second form of explainability constitutes a legal obligation at the European level, as we shall see later.

\begin{table}
 \scriptsize
  \centering
   \begin{tabular}{|l|l|l|}
     \hline
     Concept & 	Computer science & Law \\
     Decision & Partitioning of a given set of & The GDPR does not give a definition of individual \\
     ~ & data & automatized decision \\ \hline
     ~ & Sub-set: equivalence class & Profiling (art. 4 GDPR): characterized by the purpose of \\
     ~ & ~ & automated processing of personal data \\ \hline
     Automatized & Follows a pre-defined set of & GDPR: forbids automated processing which produces \\
     decision & actions without human & significant effects on persons. Included autonomous \\
     ~ & intervention & decisions (in the computer science meaning) \\ \hline
     Autonomous & General objectives defined by a & Non used concept \\
     decision & human. Learning. & ~ \\ \hline
     Explainability & Interpretability & x ante explainability : right to be informed about the \\
     ~ & Internal logic of the algorithm: & existence of an algorithmic decision \\
     ~ & relation between observations & Ex post explainability : why the algorithm has issued \\
     ~ & and outcomes & a specific decision \\
     \hline
   \end{tabular}
  \caption{Comparison of computer science and legal concepts} \label{explicability}
\end{table}

\subsection{A recent concept embodying the rule of law in the digital age}

The concerns or legal rationale for the explainability of algorithmic decisions is partly rooted in what we might call the legal meta-principle of the rule of law. At this point, it suffices to emphasize that the rule of law implies that any legal decision must, in principle, be predictable, based on a transparent process, and motivated so that it can be challenged, notably before a judge.

The rule of law aims to ensure that \emph{``Under the rule of law, all public powers always act within the constraints set out by law, in accordance with the values of democracy and fundamental rights, and under the control of independent and impartial courts. The rule of law includes principles such as legality, implying a transparent, accountable, democratic and pluralistic process for enacting laws; legal certainty; prohibiting the arbitrary exercise of executive power; effective judicial protection by independent and impartial courts, effective judicial review including respect for fundamental rights; separation of powers; and equality before the law''}\footnote{Definition from the First European Commission Report on the rule of law in the E.U.: 2020 Rule of Law Report – the rule of law situation in the European Union (COM(2020)0580), 20 September, 2020, p. 1.}. The rule of law is a cardinal legal concept. At present, respect for the rule of law is assessed on the basis of criteria that attest to the rule of law in European states both within the Council of Europe\footnote{Within the Council of Europe, the Venice Commission plays a fundamental role in the affirmation and implementation of the rule of law. Cf. its criteria list adopted in 2016 \url{https://www.venice.coe.int/webforms/documents/?pdf=CDL-AD(2016)007-f}} and the European Union.

This link between the rule of law and algorithmic law has been addressed by legal doctrine. Some authors evoke the emergence of a new form of ``algorithmic normativity'' (\cite{Barraud2018}), or even of ``algorithmic governance'' (\cite{RouvroyBerns2013}). These expressions are used both to affirm the need to regulate algorithms, especially public ones, and to note the dilution of the regulatory power of States in private actors, essentially American for the time being, or even Chinese. The right of explainability is then presented as a viaticum towards the resumption of human control over the Machine. Explainability thus has a resonance with the broader concept of digital sovereignty, which can be applied both at the individual level\footnote{Some authors use the term ``cognitive sovereignty''. See \cite{Bygrave2020}.} and a collective scale.

From the legal point of view, what matters is to align algorithmic decisions with the standards of the rule of law relative to classical legal decisions in order to maintain the democratic model. From an AI perspective, the quest goes beyond explaining how to meet the legal requirements to provide explanations for algorithmic decisions. It is a question of conducting research on the ``raison d'être'' of the use of the technical potentialities of artificial intelligence. In other words, research in machine learning has an ethical dimension: ``Being able to explain an AI-based system may help to make algorithmic decisions more satisfying and acceptable, to better control and update AI-based systems in case of failure, to build more accurate models, and to discover new knowledge directly or indirectly'' (\cite{BrkanBonnet2020}). Thus, \cite{AdadiBerrada2018} identify four main rationales for the development of XIA research: Justifying results obtained through algorithmic decisions, controlling system behavior, improving models, and increasing knowledge.

\subsection{Unstable and fuzzy normative approaches to explainability}

\subsubsection{Legal approaches: European and French law}\label{legalapproaches}

We have already previously mentioned the GDPR. It is important to point out here that it is subject to various and contradictory interpretations with regard to the explainability requirement. Some consider that the word only appears in one recital (No. 71), and thus explainability is not in the core of the Regulation (\cite{Wachteretal2017}). Others argue that it is in an overall reading, which articulates different provisions, which an explainability constraint can be found in the GDPR (\cite{BrkanBonnet2020}; \cite{Hacker2018}). Finally, for others, explainability in the GDPR is not the answer (\cite{EdwardsVeale2017}).

Apart the GDPR, the European Regulation 2019/1150\footnote{Regulation (EU) 2019/1150 of the European Parliament and of the Council of 20 June 2019 on promoting fairness and transparency for business users of online intermediation services (Text with EEA relevance).} imposes an obligation of explicability on platforms, in particular online intermediaries and search engines, with regard to their rankings These obligations are intended to be completed by other texts currently being drafted. Thus, a series of new obligations of fairness and transparency of algorithms used by digital platforms are at the heart of the adoption of the regulations on digital services legislation\footnote{Proposal for a Regulation of the European Parliament and of the Council on a Single Market For Digital Services (Digital Services Act) and amending Directive 2000/31/EC, COM/2020/825 final.} and on digital markets legislation\footnote{Proposal for a Regulation of the European Parliament and of the Council on contestable and fair markets in the digital sector (Digital Markets Act), COM/2020/842 final}. These two texts contain the seeds of new forms of algorithmic explainability around the use of algorithmic moderation tools to fight online hate, or recommendation systems.

Finally, the Artificial Intelligence Act Project\footnote{Proposal for a Regulation of the European Parliament and of the Council laying down harmonized rules on artificial intelligence (Artificial Intelligence Act) and amending certain Union legislative acts, COM/2021/206 final.} will impose explainability obligations on AI systems modulated according to risk, with reinforced obligations for high-risk AI systems. Most of these European texts currently being adopted seem to envisage above all an ex ante explanation of the use of algorithms for decision-making and the general logic of the algorithms used, so as to inform the persons concerned.

In France, the 2016 Digital Republic Act (Loi R\'epublique Num\'erique) is the starting point for French regulations on the transparency and explainability of algorithmic administrative decisions. Indeed, Article 4 of the Digital Republic Act created a new article L-311-1-1 in the Code of Relations of the Public with the Administration (CRPA), which states that: \emph{``an individual decision taken on the basis of algorithmic processing shall include an explicit statement informing the person concerned. The rules defining this processing as well as the main characteristics of its implementation shall be communicated by the administration to the person concerned if he or she so requests''}. It is mainly an obligation to communicate upon request. Article 311-3-1-2 of the CRPA specifies its scope: \emph{``The administration shall communicate to the person who is the subject of an individual decision taken on the basis of algorithmic processing, at the latter's request, in an intelligible form and subject to not infringing secrets protected by law, the following information: the degree and mode of contribution of the algorithmic processing to the decision-making; the data processed and their sources;cthe processing parameters and, where applicable, their weighting, applied to the situation of the person concerned; the operations carried out by the processing''}.

Here, explicability is essentially given a posteriori; it is not only global but also local. Beyond explicability itself, this article includes an obligation of transparency constituting a form of right to information with the obligation to explicitly mention the existence of an algorithmic decision. Moreover, \textbf{a decision taken solely on the basis of an algorithmic treatment that does not include explicit mention is considered null and void}. Etalab, a department of the interministerial digital direction, in charge of implementing the State's strategy in the field of data, has developed a guide to assist administrations in their obligations to make public algorithms explainable\footnote{ETALAB, Expliquer les algorithmes publics, \url{https://guides.etalab.gouv.fr/algorithmes/}, last version consulted 13/07/2021.}, in addition to guides on the opening of source codes\footnote{ETALAB, Ouvrir les codes sources, \url{https://guides.etalab.gouv.fr/pdf/guide-logiciels.pdf}, last version consulted 13/07/2021.}. This general obligation of explicit mention includes both exclusively automated processing operations and those that simply constitute an aid to decision-making. This compels administrations to draw up an inventory of their main algorithmic treatments. This transparency obligation is part of the broader idea that the administration must account for the use of its public algorithms. Finally, these obligations concern all individual administrative decisions, whether they concern natural persons or legal entities.

These obligations introduced in the \emph{Loi sur la République numérique} are complemented by the more specific obligations related to individual decisions based on automated processing involving the processing of personal data, as set forth in the GDPR. Thus, Article 13$\S$f) of the GDPR reinforces the right to information whenever there is an automated individual decision making. It is a matter of giving \emph{``meaningful information about the logic involved, as well as the significance and the envisaged consequences of such processing for the data subject''}. This provision seems, technically, to correspond to a principle of interpretability (\cite{Besseetal2018}). The legal doctrine sees in this at least the recognition of an overall obligation of explainability. For its part, Article 22 of the GDPR enshrines the right not to be subject to an exclusively automated individual decision, while leaving it to the Member States to authorize such algorithmic decisions, subject to certain guarantees. \textbf{It is in the context of practicing its margin of maneuver that France has adopted a specific legal basis to authorize exclusively automated individual administrative decisions}. The chosen formulation results in an extended right of explainability. Indeed, according to article 47$\S$2 of the \emph{Loi Informatique et Liberté}, the processing officer must \emph{``ensure mastery of the algorithmic processing and its developments in order to be able to explain, in detail and in an intelligible form, to the data subject the way in which the processing has been implemented with regard to him''}. This is what the doctrine calls a right of local explicability. Thus, the debate on the existence or not of a right of explicability of concrete decisions taken with respect to an individual exclusively by means of an algorithm (see \cite{Wachteretal2017}) has had little echo in French doctrine. At most, some authors have questioned the capacity of France to impose such a right of local explicability, while remaining within the limits of its margin of maneuver with respect to the GDPR (see for example \cite{Castets-Renard2018}).

In addition, the Constitutional Council has made some useful clarifications. On the one hand, it considers that ``algorithms that are likely to revise the rules they apply themselves, without the control and validation of the person in charge of the processing, cannot be used as the exclusive basis for an individual administrative decision''\footnote{Conseil Constitutionnel, Décision n° 2018-765 DC June12, 2018, point 71.}. \textbf{This interpretation seems to limit the possibility for the French administration to use so-called ``deep learning'' algorithms}. On the other hand, the Constitutional Council reminds us that the administration's duty to explain at the request of the person concerned also restricts its room for manoeuvre in the choice of the algorithmic tool. Indeed, \emph{``when the operating principles of an algorithm cannot be communicated without infringing on one of the secrets or interests set out in 2$\S$ of article L. 311-5 of the code of relations between the public and the administration, no individual decision may be taken on the sole basis of that algorithm}\footnote{Cf. Ibid., point 70.}. \textbf{In other words, the use of algorithms protected by intellectual property rights seems to be excluded in order for the administration to fulfill its obligation of transparency}. These two interpretive caveats attempt to combat the ``black box'' phenomenon, i.e., the inability to understand the exact reasons that led an algorithm to make a particular algorithmic decision. This approach illustrates France's desire to build a transparent model for algorithmic decisions that can also inspire private actors by indirectly encouraging them to implement similar safeguards.

Finally, it should be noted that explicability under Article 22 of the GDPR can only be given to the individual affected by the automated individual decision. In France, this limit was circumvented by the Constitutional Council in a decision opposing a university to a trade union that wanted to understand the logic of the algorithms used by universities to select future students. The Constitutional Council considered that it was a constitutional principle that the right of access to administrative documents allowed third parties to obtain ``in the form of a report, the criteria according to which the applications were examined and specifying, where appropriate, the extent to which algorithmic processing was used to carry out this examination''\footnote{Cf. Conseil Constitutionnel, DC- 2020-834 Union nationale des étudiants de France, 3 avril 2020, QPC, point 17. This constitutional principle was recognized on the basis of Article 15 of the Declaration of Human Rights of 1789, according to which "Society has the right to hold any public official accountable for his administration.}. This can be seen as the consecration of a French right of global explicability extended to third parties, in the case of administrative decisions based on public algorithms. This possibility opens the way to auditability by civil society.

The Law on Bioethics of August 2, 2021 is the first to introduce the notion of algorithm in connection with artificial intelligence in the public health code. The expressly provides, for the first time, an obligation of explainability for the designers of certain algorithmic treatments for health professionals who use them for an act of prevention, diagnosis or care. This article also provides for the obligation to inform the patient and to warn him or her, if necessary, of the interpretation that results from the processing of algorithmic data.

It should be noted that the government bill was enriched by parliamentary amendments. The bill only proposes a traceability of the actions of the algorithmic treatment. The deputies, inspired by the recommendations of the \emph{Commission Nationale Informatique et Liberté (CNIL)}\footnote{CNIL, Comment permettre à l’homme de garder la main ? Les enjeux éthiques des algorithmes et de l’intelligence artificielle, décembre 2017.} and the \emph{Conseil d’État}\footnote{Conseil d’État, Révision de la loi de bioéthique : quelles options pour demain? 28 juin 2018}, have proposed to add an explainability requirement, which aims to allow users of these artificial intelligence systems to understand the general logic of their operation.

This requirement implies that their designers provide users with the information necessary for this understanding and that health professionals can contribute their expertise from the moment the algorithms and health data collection strategies that feed them are developed. It also implies that healthcare professionals benefit from training that enables them to understand how these systems work in order to identify their limits and to be able to explain to patients the basis on which medical decisions concerning them are made.

\subsubsection{Why the legislator did not define explainability}

The notion of explainability is widespread in the academic world as well as in the institutional discourse at both the European and national levels\footnote{See for example, CNIL, Comment permettre à l’homme de garder la main? Les enjeux éthiques des algorithmes et de l’intelligence artificielle les enjeux éthiques, décembre 2017, p. 30, 50 ou 53; Défenseur des Droits en partenariat avec la CNIL, Algorithmes: prévenir l’automatisation des discriminations, 2020 p. 8; Commission Nationale consultative des droits de l’Homme, Avis sur la lutte contre la haine en ligne, 8 juillet 2021, p. 27.}. The lack of use of the term explainability by legislators is justified because the concept is so loose and has only been introduced into law in contextualized form. \cite{Maxwell2020} rightly states that explainability depends on four important factors: \\
 - The recipient of the explainability, i.e. the audience targeted by the explanation. For example, its level will be different depending on whether it is a user or a regulator. \\
 - The level of importance and impact of the algorithm. The explainability of an autonomous car accident does not have the same level of importance as that of an advertising or video recommendation algorithm. \\
 - The legal and regulatory framework, which is different in different geographical areas, such as in Europe with the General Data Protection Regulation (GDPR). \\
 - The operational environment of explainability, such as its mandatory nature for certain critical applications, the need for certification prior to deployment, or the facilitation of use by users (\cite{Maxwell2020}, p. 14).

Thus, there is not a general right to explainability in digital law, but rather obligations of explainability introduced in various regulations, as we have already mentioned. It should be remembered that most of these require the intervention of the regulatory power, supplemented by regulatory authorities that try to explain how to implement this right of explainability. However, this notion is most often linked to a requirement for transparency.

\subsubsection{A recent concept most often associated with the principle of transparency}

\noindent The High-Level Expert Group on Artificial Intelligence (AI HLEG)\footnote{The High-Level Expert Group on Artificial Intelligence (AI HLEG): The Ethics Guidelines for Trustworthy Artificial Intelligence (AI), document prepared by the AI HLEG for the European Commission, April, 2019.} enshrines explainability as one of the four ethical principles that any AI system must respect, alongside respect for human autonomy, prevention of harm to human beings and fairness\footnote{The AI HLEG considers that “Explicability and Responsibility are closely linked to the rights relating to Justice” (as reflected in Article 47 of the EU Charter of fundamental rights).}. Generally speaking, the HLEG links explainability to transparency, alongside the principles of traceability and communication. The HLEG considers that: ``Explicability is crucial for building and maintaining users' trust in AI systems. This means that processes need to be transparent, the capabilities and purpose of AI systems openly communicated, and decisions – to the extent possible – explainable to those directly and indirectly affected''.

Similarly with the OECD recommendation on AI, transparency and explainability are the third of the five principles that underpin a responsible approach to AI that can generate trust. The other principles are i) inclusive growth, sustainable development and well-being; ii) human-centred values and fairness, iii) transparency and explainability; iv) robustness, security and safety; and v) accountability. In paragraph 1.3. on ``transparency and explainability'', the OECD considers that: \\
\emph{``AI Actors should commit to transparency and responsible disclosure regarding AI systems. To this end, they should provide meaningful information, appropriate to the context, and consistent with the state of art: \\
i.to foster a general understanding of AI systems, \\
ii.to make stakeholders aware of their interactions with AI systems, including in the workplace, \\
iii.to enable those affected by an AI system to understand the outcome, and, \\
iv.to enable those adversely affected by an AI system to challenge its outcome based on plain and easy-to-understand information on the factors, and the logic that served as the basis for the prediction, recommendation or decision.''}

Transparency is conceived as a broader principle that can be satisfied with making raw information accessible or intelligible, whereas explainability seems to imply the need to make the information given understandable\footnote{\cite{Beaudouinetal2020a}, p. 10 contrast with other authors who consider that transparency is a sub-principle of explicability. }. Thus, the opening of the source code of an algorithm is most often linked by jurists to the principle of transparency and more indirectly to explainability because this code is rarely understandable to a layman. From the outset, the principle of explainability raises a difficulty of demarcation with respect to other concepts, which explains why its field of application is debated.

\subsubsection{A concept whose scope is debated}

The principle of explainability raises four main types of debate about its scope, which are closely intertwined.

\textbf{Firstly}, it is important to note that constructing explanations, justifications, interpretations is not a straightforward and neutral activity based upon exclusively technical specifications. It is purposeful and subjectively established following three dimensions (\cite{Barale2021}).

\begin{enumerate}
  \item \textbf{Explanations for whom?} Algorithms, platforms and other autonomous artifacts are designed, coded, used by and produce impacts for several different stakeholders: software engineers, indirect users (who specify the tool, but do not directly use it), end users (the ones who actually use it for some specific purpose), impacted citizens/customers/clients, the society as a whole, etc. Each of them has different types of expectations from such artifacts and consequently expects different types of explanations.
  \item \textbf{Explanations for which need?} Given the diversity of implied stakeholders it is natural to expect a diversity of motivations and specifications for explanations. Such generic motivations include testing, checking, monitoring, understanding and appealing/revising/updating.
  \item \textbf{Explanations for doing what?} Besides explanations designed for some generic purpose there are specific uses of the explanations which need to be considered and anticipated (as much as possible). Under such a perspective, explanations constructed for some specific purposes generally do not fit the requirements of other purposes and other specific stakeholders who might want to use explanations.
\end{enumerate}
	
\textbf{Secondly}, this semantic debate is in fact a debate about the object of explainability. The duality of the object of explainability has been recalled by the HLEG, which distinguishes between two types of explainability. On the one hand, technical explainability, and on the other hand, explainability linked to human decisions taken on the basis of this technique:

\emph{``Technical explainability requires that the decisions made by an AI system can be understood and traced by human beings. Moreover, trade-offs might have to be made between enhancing a system's explainability (which may reduce its accuracy) or increasing its accuracy (at the cost of explainability). Whenever an AI system has a significant impact on people's lives, it should be possible to demand a suitable explanation of the AI system's decision-making process. Such explanation should be timely and adapted to the expertise of the stakeholder concerned (e.g. layperson, regulator or researcher). In addition, explanations of the degree to which an AI system influences and shapes the organisational decision-making process, design choices of the system, and the rationale for deploying it, should be available (hence ensuring business model transparency)''}\footnote{HLEG, Ethics Guidelines for Trustworthy AI, 8 April 2019, p. 18.}.

Thus, the contextual character of explainability necessarily implies that its object must be precisely determined. Schematically, explainability can concern the choice of a data set, the algorithm that will use the data, the model that will be generated and finally the content, the decision or the prediction that will be made on the basis of this model. Even when explainability only applies to the algorithm, some authors try to classify the possible forms of explainability according to the type of algorithm involved.

\textbf{Thirdly}, there is a debate about the extent of explainability. This debate has several aspects. On the one hand, there is a debate on the necessity, the relevance or not of systematically providing for a form of explainability. Some authors propose a classification of artificial intelligence systems according to risk, with the objective of limiting explainability to certain types of high-risk artificial intelligence (\cite{Robbins2019}). This is the direction in which the European legislator seems to want to go. Article 13 of the recent European proposal for a regulation limits transparency and the provision of information to users to high-risk AI systems. Beyond the risk approach, some authors consider identifying the degree of explainability in a given situation through an analysis of costs and benefits for society (\cite{Beaudouinetal2020a}). The same authors identified seven categories of costs to be considered in this evaluation: design, reduction of prediction accuracy, creation and storage of logs, violation of business secrecy, conflict with security and other policy objectives, reduction of decision flexibility in the future, and slowing of innovation. As for the operational benefits, they refer to the ability to foster user confidence in the system, and the ability to make the algorithms more robust and certifiable. These different criteria must be taken into consideration in a contextual manner.

The extent of explainability is also debated in relation to its recipient: expert, regulator or individual. Indeed, the intelligibility of the explanations given necessarily varies according to the knowledge of the recipient. Very detailed and complex information will not be very useful for the layman, who will mainly be looking to know in simple terms what algorithmic model was used and what criteria were used in the decision taken against him.

The scope of explainability is central to other related notions, including traceability and audibility. As the HLEG outlines, ``An explanation as to why a model has generated a particular output or decision (and what combination of input factors contributed to that) is not always possible. These cases are referred to as ``black box'' algorithms and require special attention. In those circumstances, other explicability measures (e.g. traceability, auditability and transparent communication on system capabilities) may be required, provided that the system as a whole respects fundamental rights'' (HLEG, Ethics guidelines for trustworthy AI, 2019,  p. 13).

Thus, traceability can be a minimum requirement when full explainability is not possible. In this context the HLEG defines traceability as “The data sets and the processes that yield the AI system’s decision, including those of data gathering and data labelling as well as the algorithms used, should be documented to the best possible standard to allow for traceability and an increase in transparency. This also applies to the decisions made by the AI system. This enables identification of the reasons why an AI-decision was erroneous which, in turn, could help prevent future mistakes” (HLEG, Ethics guidelines for trustworthy AI, 2019, p. 18).

Traceability can be based on logging obligations, i.e., memorizing the various actions carried out by a model or an algorithm. Auditability goes further since it allows a third party to check ex ante and/or ex post the model and/or the algorithmic decisions taken on the basis of a model. Auditability is therefore a means of verifying the relevance of the explanations provided in the context of both global and local explainability. Although distinct, the notions of traceability and auditability are closely
linked to explainability.

\textbf{Fourthly}, there is a debate about the limits to the right of explainability. This debate is prominent among lawyers who are concerned with the extent to which explainability could be challenged on the grounds of intellectual property rights, business secrets, or other secrets related to defense or national security (\cite{BrkanBonnet2020}; \cite{Maggiolino2019}).
In machine learning research, it is mainly the technical limits of explainability that are examined. It is sometimes necessary to explain that in certain situations explanations can only be partial or should only focus on the most important reasons that led to the decision in a specific situation. At this point, it is worth mentioning the existence of a bi-disciplinary research (combining machine learning and law) whose purpose is to come up with a grid for assessing the technical feasibility of the explainability obligation enshrined in Article 22 of the GDPR (\cite{BrkanBonnet2020} – table \ref{picture4}).

\begin{table}
  \centering
  \includegraphics[scale=0.85]{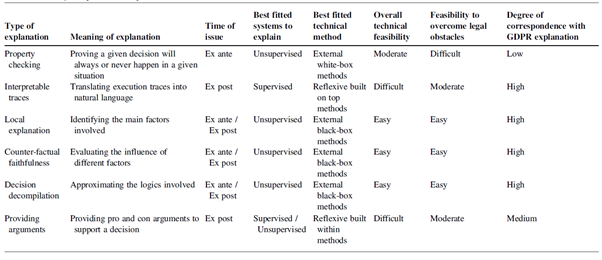}
  \caption{Practical feasibility of explanation of algorithmic decisions (source: \cite{BrkanBonnet2020} p. 47)}\label{picture4}
\end{table}

Assessing the technical feasibility of the explainability requirement is interesting, but while Article 22 of the GDPR is important, it is not the only legal provision that addresses the topic. In addition, the basis and method for qualifying the technical feasibility of different types of explanations are not detailed by the authors. We now come to the observation that there is a certain diversity, not to say heterogeneity, in the legal provisions on explainability in European Union law and in French law.

\subsubsection{Heterogeneity of legal provisions and purposes in the European and French law}

The GDPR includes, in addition to Article 22, two important provisions: Article 15 and Recital 71. The two articles have different objectives: the information obligation (Art. 15) and the right to a final human decision (Art. 22); they give rights to individuals, albeit with restrictions for beneficiaries of the right to a humane decision under Article 22 (see supra subsection \ref{legalapproaches} and Table \ref{explainEU}, Appendix 1 below). On the other hand, Recital 71, as well as Article 29 of the Working Party Guidelines (which inspired the GDPR), target algorithm experts, designers and programmers. These two texts have the objective of improving models, while the previous ones have other purposes. Finally, the European Regulation 2019/1150 already mentioned, targets economic actors (providers of online intermediation services or search engines) and requires them to be in a position to explain (to the regulator or the judge) the ranking parameters used in their algorithms.

In French law, two laws (the Computer and Liberty Act and the Bioethics Act), as well as the Code of Relations of the Public with the Administration and two decisions of the Constitutional Council, provide a contrasting landscape of rules relating to the explainability of algorithmic decisions (Table \ref{explainFR}, Appendix 2 below).

In its decision of June 12, 2018, the Constitutional Council, seized by 60 senators against the law transposing European texts on personal data, decided that the administration must refrain from using autonomous algorithms and could not invoke the protection of algorithms by intellectual property rights or commercial secrecy to exempt itself from an obligation of explainability of algorithmic decisions. In its decision of April 3, 2020, seized of a priority question of constitutionality submitted by the Conseil d’Etat confronted with the appeals of a student union (UNEF) requesting that universities make public the information relating to the criteria used, including by algorithmic treatments, to decide to admit or refuse a university application (via the Parcours Sup platform). On the other hand, the Code of Relations of the Public with the Administration and the Bioethics Law define explainability in terms of information about the structure of models, but the former concerns individuals while the latter also concerns healthcare professionals and designers of algorithms used in healthcare.

\section{Conclusion}

The central argument advanced in this article is that achieving fairness in machine learning algorithms cannot be handled by isolated disciplines: legal science and algorithmic technique, in its mathematical and computational components. Like others (\cite{AbuElyounes2020}; \cite{Besseetal2018}; \cite{Hacker2018}; \cite{Wachteretal2020}, \cite{Wachteretal2021}; \cite{Xiang2021}) we argue that the two disciplines must cross and enrich their perspectives. On the one hand, the design of the models and the tests used must be understood by social scientists and lawyers; on the other hand, the views proposed by the latter, combined with the understanding of legal mechanisms and their cultural and institutional context, can be usefully integrated by AI researchers and machine learning algorithm designers.
Though interdisciplinary cooperation is not simple, it is desirable that it be developed both to produce knowledge and to provide guidance for public decision-making. Researchers could promote work on fairness based on society's prevailing values and principles, which amounts to instrumentalizing the modeling process on the basis of social preferences embodied in legislation  and case-law. It is also useful to take a critical look at the latter in order to improve their relevance, if needed. Thus, recommendations could be made to the legislator, the regulator or the judge. We have argued that it would be useful to better legislate on the explainability of algorithmic decisions, by imposing clear and realistic objectives. Interdisciplinary research paths can be outlined.

\begin{enumerate}
  \item A division of labor between computer scientists and social scientists was mentioned earlier, when \cite{Wachteretal2021} claimed that law and public policy alone have the capacity to shift historical social biases toward greater equity (the notion of “bias transforming”), while the search for "fair machine learning" would be fatally condemned to use "bias preserving metrics". This division of labor arguably does not do justice to the vigor of machine-learning research that aims to make the world a better place via nondiscriminatory algorithms (\cite{Feldmanetal2015}), but which can be made more effective with the cooperation of lawyers and other social scientists.
  \item The study of preferences encapsulated in algorithms would be worth further exploration. As \cite{Binns2018} argues, fairness in machine learning can refer to a variety of egalitarian considerations, and thus to criteria and theories of social justice, which it is legitimate to explain and justify. In a similar vein, \cite{Tsoukias2021} argues that there are preferences and values in any decision support system; therefore: \emph{``If an autonomous artefact is able to make a decision or to compute a recommendation, it means that somebody embedded within the artefact his/her preferences. And these are independent from how the artefact turns to learn out from the data feeding it. It turns out that is of paramount importance to know how values are actually embedded in any of such systems and/or how these are learned''} (\cite{Tsoukias2021}, p. 158). \cite{Friedleretal2016} analyze the variables, unobservable but significant for the prediction, which must be taken into account as an intermediate space between the space of inputs and that of outcomes. They then show that the axiomatics of the decision-maker's choice (for the selection of candidates for university entrance) have implications on the outcomes and their (in)fairness: the decision-maker can estimate that there is no distance between the space of unobservable variables and that of observed variables (axiom: ``What you see is what you get'') or that these spaces are separate; in this case, the decision-maker can postulate that the axiom ``we're all equal'' and, thus, believe that variables that explain different educational performances by school location should not be considered in the selection process. The important thing here is to emphasize that the choices made on the basis of algorithms are based on principles (an axiomatic) laid down by the decision-maker. This leads us to a third proposal.
  \item By generalizing the previous considerations, we arrive at the problem of the social choice of a fairness norm. Does society prefer to seek equality of opportunity, or demographic parity, (multi)calibration, causality, or any other form of fairness? A fundamental question also arises: is fairness the only social value sought? is it compatible with other social objectives? In this regard, for example \cite{Corbett-Davisetal2017} show, in the case of COMPAS recidivism risk and in utilitarian terms, that pursuing the goal of maximizing public safety is not necessarily compatible with the equal treatment of individuals across race: \emph{``Since the optimal constrained and unconstrained algorithms in general differ, there is tension between reducing racial disparities and improving public safety} (\cite{Corbett-Davisetal2017}). Moreover, given that social choices can be differentiated according to domains, public policies and law can convey several logics (for example, between penal and public security policies and social protection or access to university). Modeling composite social choices could inform public debates and legislators, who could be better informed about the implications of choices made\footnote{For example, computer scientists can show the impossibilities, such as the reconciliation between individual and group non-discrimination (\cite{Friedleretal2016}.}. Furthermore, can the possibility of a gap between the social values held by the public decision-maker and those of civil society be modelled in terms of bi-sided fairness?
  \item The problem of the explainability of automatic algorithmic decisions deserves to be further investigated. Is it possible to conceive a method or methods of technical evaluation of the feasibility of explainability as imposed by the law, with its zones of vagueness or indeterminacy? Is it possible to demonstrate that explainability can be a way to achieve fairness in algorithmic decisions?
  \item While work on anti-discrimination law exists in the United States and the European Union, very little research has been done on national legal and institutional systems. A better knowledge of the legal and political stakes of national systems (in Europe, Asia, Oceania, Latin America) could feed the multidisciplinary research avenues to which this article is addressed.
\end{enumerate}

\section*{Acknowledgements}

The paper is part of the interdisciplinary projet INTERFAIR, funded by the CNRS MITI (\emph{Mission pour l'Interdisciplinarité}). The last author acknowledges the many discussions with the colleagues of DIMACS (Rutgers, USA) and the 3A Institute (ANU, AU) and all the participants within the ``Social Responsibility of Algorithms'' workshops (\url{https://www.lamsade.dauphine.fr/sra}).

\bibliographystyle{plain}
\bibliography{interfair}

\newpage


\begin{table}[b]
\tiny
  \centering
   \begin{tabular}{|l|l|l|l|l|l|}
   \multicolumn{6}{l}{\normalsize{\textbf{Appendix 1}}} \\
   \multicolumn{6}{l}{\normalsize{\textbf{~}}} \\
   \multicolumn{6}{l}{\normalsize{\textbf{~}}} \\
   \multicolumn{6}{l}{\normalsize{\textbf{~}}} \\
   \multicolumn{6}{l}{\normalsize{\textbf{~}}} \\
   \multicolumn{6}{l}{\normalsize{\textbf{~}}} \\
     \hline
     \textbf{E.U. legal text} & \textbf{Disposition} & \textbf{Scope of application} & \textbf{Practical modalities.} & \textbf{Who is concerned} & \textbf{Goals} \\ \hline
     Article 29 & ``algorithmic auditing'': & ~ & ~ & ~ & ~ \\
     Working & ``testing the algorithms & ~ & ~ & ~ & ~ \\
     Party & used and developed by & ~ & ~& ~ & ~ \\
     Guidelines on & machine learning systems& ~ & Formal & ~ & ~ \\
     Automated & to prove that they are & ~ & verification of & Experts: & Model \\
     individual & actually performing & Without restriction & algorithms + & Designers & improvement \\
     decision- & as intended and not & ~ & security & programmers & ~ \\
     making and & producing erroneous& ~ & ~ & + certifiers & ~ \\
     Profiling. & discriminatory or & ~ & ~ & ~ & ~ \\
     ~ & unjustified results & ~ & ~ & ~ & ~ \\ \hline
     ~ & Article 15(1)(h): right to & ~ & ~ & ~ & ~ \\
     GDPR & be informed of the & ~ & ~ & ~ & ~ \\
     ~ & existence of automated & ~ & ~ & ~ & ~ \\
     ~ & decision-making + logic & ~ & Internal logic & ~ & ~ \\
     ~ & involved + significance & Without restriction & of the model & Individuals & Transparency \\
     ~ & and the consequences & ~ & + causality & ~ & + causality \\
     ~ & of such processing & ~ & ~ & ~ & ~ \\
     ~ & for the data subject & ~ & ~ & ~ & ~ \\ \hline
     ~ & Article 22 : right not to & Not applicable if & Concerns & ~ & ~ \\
     ~ & be subject to a decision & - entering into, or & both & ~ & ~ \\
     ~ & based solely on & performance of, a & automated & ~ & ~ \\
     ~ & automated processing, & contract & decision & ~ & Final decision \\
     ~ & including profiling, which & - authorised by Union & (explainable) + & Individuals & must be taken \\
     ~ & produces legal effects & or Member State law & autonomous & ~ & by a human \\
     ~ & concerning him or her or & - no legal effects or & decision & ~ & ~ \\
     ~ & similarly significantly & person not affected & (black box, & ~ & ~ \\
     ~ & affects him or her & - consent, legal & not explainable) & ~ & ~ \\
     ~ & ~ & authorization & Causality & ~ & ~ \\ \hline
     ~ & Recital n° 71: & ~ & ~ & ~ & ~ \\
     ~ & use of appropriate & ~ & ~ & ~ & ~ \\
     ~ & mathematical or & ~ & ~ & ~ & ~ \\
     ~ & statistical procedures & ~ & ~ & ~ & ~ \\
     ~ & implement technical & ~ & ~ & ~ & ~ \\
     ~ & and organisational & ~ & Pipe: & ~ & ~ \\
     ~ & measures ... & ~ & Formal & Experts: & Model \\
     ~ & ensure that factors & Without restriction & verification & Designers & Improvement \\
     ~ & which result in & ~ & Data control & Programmers & ~ \\
     ~ & inaccuracies in personal & ~ & Test & ~ & ~ \\
     ~ & data are corrected and & ~ & ~ & ~ & ~ \\
     ~ & the risk of errors is & ~ & ~ & ~ & ~ \\
     ~ & minimized and prevents & ~ & ~ & ~ & ~ \\
     ~ & discriminatory effects & ~ & ~ & ~ & ~ \\ \hline
     Regulation (EU) & Article 5 – online & ~ & ~ & Industry: online & ~ \\
     2019/1150 & intermediation services - & Without restriction	& Verification & intermediation & Transparency \\
     ~ & main parameters & ~ & ranking model & service providers  & + compliance \\
     ~ & determining ranking & ~ & ~ & + search engines & ~ \\ \hline
   \end{tabular}
  \caption{Explainability in European Union Law}\label{explainEU}
\end{table}

\newpage

\begin{table}[b]
\tiny
  \centering
   \begin{tabular}{|l|l|l|l|l|l|}
   \multicolumn{6}{l}{\normalsize{\textbf{Appendix 2}}} \\
   \multicolumn{6}{l}{\normalsize{\textbf{~}}} \\
   \multicolumn{6}{l}{\normalsize{\textbf{~}}} \\
   \multicolumn{6}{l}{\normalsize{\textbf{~}}} \\
   \multicolumn{6}{l}{\normalsize{\textbf{~}}} \\
   \multicolumn{6}{l}{\normalsize{\textbf{~}}} \\
     \hline
     \textbf{E.U. legal text} & \textbf{Disposition} & \textbf{Scope of application} & \textbf{Practical modalities.} & \textbf{Who is concerned} & \textbf{Goals} \\ \hline
     Code of Relations & Right to information upon & Décisions algorithmiques de & Structure & Individuals	& Information \\
     of the Public with & request of the person concerned & l’administration & du modèle & ~ & + transparency \\
     the Administration & ~ & (organisation privées exclues) & ~ & ~ & ~ \\
     (incorporation of Loi & ~ & ~ & ~ & ~ & ~ \\
     sur la République & ~ & ~ & ~ & ~ & ~ \\
     numérique) & ~ & ~ & ~ & ~ & ~ \\ \hline
     Computer and & Art. 47$\S$2: explanation in detail & Without restriction & Pipe: & Individuals	& Justification \\
     Freedom Law & and in an understandable form: & ~ & Formal & ~ & of the \\
     (Loi Informatique & - contribution to decision making & ~ & verification & ~ & decision \\
     et Liberté) & - data & ~ & Data & ~ & ~ \\
     ~ & - processing parameters of the & ~ & control & ~ & ~ \\
     ~ & operations carried out & ~ & Test + & ~ & ~ \\
     ~ & ~ & ~ & Causality & ~ & ~ \\ \hline
    Conseil & Restriction on the use of deep & Algorithmic decisions of the & Not & Individuals & Limiting \\
    constitutionnel & learning (autonomous & administration (excluding & relevant & ~ & autonomous \\
    (June 12, 2018 & algorithms) and algorithms & private organizations) & ~ & ~ & decisions \\
    decision on the & protected by IPR or secrecy & ~ & ~ & ~ & ~ \\
    personal data law) & ~ & ~ & ~ & ~ & ~ \\ \hline
    Decision of Conseil & Right of third parties to & Selection algorithm for entrance & Not & Third party & Transparency \\
    constitutionnel & explanation (students' union & to the University (Parcours Sup) & relevant & (non expert) & ~ \\
    (April 3, 2020) & Union Nationale des Etudiants & ~ & ~ & ~ & ~ \\
     - Communicability & de France - UNEF,) & ~ & ~ & ~ & ~ \\
    and publicity of & ~ & ~ & ~ & ~ & ~ \\
    university & ~ & ~ & ~ & ~ & ~ \\
    admission & ~ & ~ & ~ & ~ & ~ \\
    algorithms & ~ & ~ & ~ & ~ & ~ \\ \hline
    Law on bioethics & Article 17 – Information on the & Algorithmic decisions & Structure & Individuals & Information \\
    of August 2, 2021 & use of a learning algorithm (for & in the health sector & of the model & Health & + transparency \\
    ~ & prevention, diagnosis or care) & ~ & ~ & professionals & ~ \\
    ~ & and the resulting interpretation; & ~ & ~ & Experts: & ~ \\
    ~ & the designers of the algorithmic & ~ & ~ & designers & ~ \\
    ~ & processing ensure explainability & ~ & ~ & ~ & ~ \\
    ~ & of its operation for users & ~ & ~ & ~ & ~ \\ \hline
   \end{tabular}
  \caption{Explainability in French Law}\label{explainFR}
\end{table}

\end{document}